\documentclass[%
reprint,                
 amsmath,amssymb,
 aps,
 prx,
]{revtex4-2}

\usepackage{graphicx}
\usepackage{dcolumn}
\usepackage{bm}

\usepackage{float} 
\usepackage{tikz}
\usetikzlibrary{positioning}
\usetikzlibrary{calc}
\DeclareMathOperator{\sech}{sech}

\usepackage{amsmath,amssymb,mathtools} 
\usepackage{float}                     
\usepackage[ruled,vlined,linesnumbered]{algorithm2e}
\DontPrintSemicolon
\SetKwInOut{Input}{Input}
\SetKwInOut{Output}{Output}
\SetKwComment{tcp}{\hspace{0.5em}// }{}

\usepackage{amsthm}

\theoremstyle{remark}

\usepackage{booktabs} 

\begin{document}

\preprint{APS/123-QED}

\title[Reshaping Global Loop Structure to \\ Accelerate Local Optimization by Smoothing Rugged Landscapes]{Reshaping Global Loop Structure to \\ Accelerate Local Optimization by Smoothing Rugged Landscapes}

\author{Timothée Leleu}
\email{timothee.leleu@ntt-research.com,tleleu@stanford.edu}
\affiliation{NTT Research, Stanford University}

\author{Sam Reifenstein}
\affiliation{NTT Research, UC Berkeley}

\author{Atsushi Yamamura}
\affiliation{Stanford University}

\author{Surya Ganguli}
\affiliation{Stanford University}

\date{\today}

\begin{abstract}

Probabilistic graphical models with frustration exhibit rugged energy landscapes that trap iterative optimization dynamics. These landscapes are shaped not only by local interactions, but crucially also by the global loop structure of the graph. The famous Bethe approximation treats the graph as a tree, effectively ignoring global structure, thereby limiting its effectiveness for optimization. Loop expansions capture such global structure in principle, but are often impractical due to combinatorial explosion. The $M$-layer construction provides an alternative: make $M$ copies of the graph and reconnect edges between them uniformly at random. This provides a controlled sequence of approximations from the original graph at $M=1$, to the Bethe approximation as $M \rightarrow \infty$. Here we generalize this construction by replacing uniform random rewiring with a structured mixing kernel $Q$ that sets the probability that any two layers are interconnected. As a result, the global loop structure can be shaped without modifying local interactions. We show that, after this copy-and-reconnect transformation, there exists a regime in which layer-to-layer fluctuations decay, increasing the probability of reaching the global minimum of the energy function of the original graph (equivalently, the maximum-{\it a-posteriori} configuration). This yields a highly general and practical tool for optimization in problems with complex global loop structure. Using this approach, the computational cost required to reach these optimal solutions is reduced across sparse and dense Ising benchmarks, including spin glasses and planted instances. When combined with replica-exchange Monte Carlo, the same construction increases the polynomial-time algorithmic threshold for the maximum independent set problem. A cavity analysis shows that structured inter-layer coupling significantly smooths rugged energy landscapes by collapsing configurational complexity and suppressing many suboptimal metastable states. In addition to landscape smoothing, our analysis identifies an emergent Nesterov-like acceleration in the optimization dynamics induced by inter-layer interactions, arising naturally from drift in a one-dimensional ring mixing across layers.

\end{abstract}

\maketitle


\clearpage

\section{Introduction}

Energy landscapes arising in disordered systems, spin glasses, and combinatorial optimization are often rugged, causing local-update algorithms to become trapped far from optimal configurations. Many optimization problems become hard not only in the worst case, but also on typical instances, when frustrated loops arise in the interaction graph~\cite{mezard2009information}. In replica theory, it is possible to account for the number of metastable states that impede local algorithms~\cite{parisi1979toward,mezard1987spin,Monasson1995}. Locally tree-like graphs form a special class that is well understood analytically. In the limit where loops in the interaction graph are long and rare, cavity theory~\cite{pearl1988probabilistic,mezard2002random,mezard2003cavity,montanari2015finding,zdeborova2007phase} yields the belief-propagation (BP) equations, whose fixed points describe stationary points of the Bethe free energy. Essentially, the cavity theory provides a variational framework for describing marginal distributions in the tree-like (Bethe) approximation, with replica-symmetry breaking (RSB) accounting for the proliferation of metastable states.

Nevertheless, this simpler picture breaks down when many short loops are present. Dense graphs can often be handled by other variational schemes instead \cite{montanari2021optimization,sarao2020marvels}. For graphs in the intermediate regime, it is harder to define a useful simplifying ansatz. Loop corrections to belief propagation have been derived \cite{chertkov2006loop,montanari2005compute}, yet they are practical only for small loops and small systems \cite{kirkley2021belief}. Many variants of BP have been proposed to push beyond the tree limit \cite{marino2016backtracking}, but none gives a general cure for many short loops.

A more systematic route to organizing loop corrections is provided by the $M$-layer construction~\cite{lucibello2014finite,altieri2017loop,angelini2020loop,angelini2022unexpected,angelini2025bethe}, which in its standard form makes $M$ copies of the base graph and permutes the connections between them uniformly at random. One thereby obtains a larger ``lifted'' graph in which the local couplings remain unchanged, while short loops in the original graph are stretched across different layers.  As $M$ increases, the lifted graph becomes effectively tree-like. In the limit $M \to \infty$, the free energy of the lifted graph converges to the Bethe free energy of the original graph\cite{vontobel2013counting}, while for finite $M$ the residual loops give controlled corrections that can be organized as a diagrammatic expansion, closely analogous to loop expansions in field theory \cite{lucibello2014finite,altieri2017loop}.

So far, the $M$-layer construction has served mainly as a theoretical tool for studying critical behavior in lattice models \cite{altieri2017loop,angelini2020loop,angelini2022unexpected,angelini2025bethe}, not as a method for optimization. Existing versions effectively describe a statistical mixture of Bethe states and offer no mechanism for selecting or decimating toward a single solution \cite{braunstein2005survey}.

The purpose of this work is to develop an $M$-layer graph-lifting construction applicable to optimization that allows explicit control over the \emph{global} pattern of correlations induced by loops through a tunable layer-to-layer connectivity, while preserving local interaction structure.

Our construction begins with a factor graph and copies every node $M$ times. Connections between copies (or layers) are then rewired using a biased random matching: a tunable mixing matrix $Q$ specifies how frequently edges originating in one layer connect to another. Uniform $Q$ recovers the standard $M$-layer construction with uniformly random rewiring across layers, while structured choices of $Q$ bias the rewiring toward specific layer-to-layer connections and shape how information propagates across the lifted system. As a result, information flows in two coupled directions: across the original interaction graph and along the inter-layer network defined by $Q$. While many layer-interaction graphs are possible in principle, we focus in the following on the simple but effective case of a Gaussian-drift ring mixer.
\\
\\
For context, approaches based on coupling multiple copies of a system have also been explored for optimization, most notably in replicated simulated annealing (RSA) \cite{baldassi2016unreasonable,angelini2019monte}. In this approach, copies of the base problem are coupled through an explicit ferromagnetic interaction that encourages corresponding variables across layers to align. 
Such replica coupling has been observed to effectively smooth the energy landscape in certain settings \cite{baldassi2016unreasonable}. However, RSA modifies the local neighborhood of each spin by introducing additional inter-replica interactions, in contrast to the $M$-layer construction which preserves local interactions; moreover, in existing formulations this ferromagnetic inter-replica coupling is typically homogeneous and does not impose an explicit structure on how different replicas are interconnected. In harder combinatorial problems such as the maximum independent set, its performance gains have been shown to be limited, with only modest improvements over parallel tempering reported \cite{angelini2019monte}. At the same time, RSA can yield substantial improvements over single-replica baselines in other settings, such as planted graph coloring \cite{angelini2023limits}.

Ideas akin to cloning a graph have also appeared in coding theory: spatially coupled low-density parity-check codes (LDPC) replicate and link code blocks to approach Shannon-limit decoding \cite{kudekar2011threshold,kumar2014threshold,yedla2014simple,aref2015approaching,ren2024towards,hamed2013threshold,macris2012beyond,olmos2017continuous,liu2023finite}. These methods rely on coupling along a chain, whereas here the goal is to design general graph-lifting operations that improve convergence of local dynamics.
\\
\\
In this paper, we first empirically characterize the effect of the structured $M$-layer lift across a broad range of Ising optimization problems, from sparse random regular graphs to dense mean-field models. We find that the residual energy reached by greedy relaxation decreases with the number of layers $M$. Although each sweep on the lifted graph is more computationally expensive, the probability of reaching the ground state increases sufficiently that the total computational cost is reduced for both greedy dynamics and simulated annealing. When combined with state-of-the-art Monte-Carlo–exchange schemes, the structured lift raises the algorithmic threshold for the maximum independent set problem, i.e., the largest independent-set density reachable in polynomial time, compared to parallel tempering \cite{swendsen1986replica,hukushima1996exchange} (PT) and replicated simulated annealing \cite{baldassi2016unreasonable,angelini2019monte} (RSA).

To explain these empirical observations, we develop a cavity-theoretic analysis of the structured $M$-layer. We derive belief-propagation equations with mixing and show that, to leading order, the lifted dynamics corresponds to stochastic descent on the Bethe free energy of the base graph, with a noise amplitude controlled by coherent fluctuations across layers and exhibiting a Nesterov-like acceleration when the ring carries a drift. Extending the $M$-layer cavity theory to the 1-RSB level, we compute the configurational complexity and show that it decreases with increasing $M$, indicating a smoothing of the rugged landscape of metastable states, consistent with the observed improvement in optimization performance.

The structured $M$-layer lift is completely general. Because it modifies only the interaction topology and not the update rule, it can be combined with many iterative algorithms. It applies to arbitrary probabilistic graphical models and is therefore compatible with combinatorial optimization, probabilistic decoding, and related machine-learning tasks. More broadly, this framework opens the possibility of designing richer layer-to-layer interaction graphs beyond the simple 1D ring example used here, allowing one to exploit instance-specific structure and potentially achieve even greater speedups.
\\
\\
Section \ref{sec:method} defines the structured $M$-layer transform. Section \ref{sec:empirical} shows how this construction improves MAP inference through numerical experiments on various benchmarks. Section \ref{sec:mlayercavity} develops the cavity-theory description, deriving the message-passing equations with mixing and the corresponding contraction criterion. Finally, Section \ref{sec:mlayer1RSB} extends the analysis to the 1-RSB level, where the collapse of inter-layer variability appears as a measurable reduction in landscape ruggedness, as measured by configurational complexity.

\section{Method \label{sec:method}}

\subsection{Factorizable probabilistic model}

The structured M-layer transformation applies to any generic factorizable probabilistic model defined on a factor graph $G=(V,A)$, where $V$ denotes a set of variable nodes, and $A$ denotes a set of interaction factors. Here we will restrict to variable nodes $i \in V$ that take values $x_i$ in a finite alphabet but the transformation can be easily generalized to the continuous case. The factor graph is described by its unary factors $f_i(x_i)$ for each variable $i \in V$, and interaction factors $f_a(x_{\partial a})$ for each factor $a \in A$, where $\partial a \subseteq V$ denotes the variables involved in factor $a$. The Gibbs distribution of the model is
\begin{align}
P(x) = \frac{1}{Z} \prod_{i \in V} f_i(x_i) \prod_{a \in A} f_a(x_{\partial a}),
\label{eq:Gibbs}
\end{align}
\noindent where $Z$ is the partition function given as $Z = \sum_{\{x\}} \prod_{i \in V} f_i(x_i) \prod_{a \in A} f_a(x_{\partial a})$. Each factor \( f_a(x_{\partial a} \mid \theta_a) \) depends on instance-specific disorder \( \theta_a \). For synthetic benchmarks, these parameters are typically drawn from a prescribed distribution \( P_\theta \). The Ising model with Hamiltonian $H(x) = - \sum_{i<j} J_{ij} x_i x_j - \sum_i h_i x_i$ and $x_i \in \{-1,1\}$ is recovered as the special case with $f_i(x_i) = e^{\beta h_i x_i}$ and $f_{(ij)}(x_i, x_j) = e^{\beta J_{ij} x_i x_j}$, where each instance is specified by its parameters $\theta=(J,h)$ with couplings \(J_{ij}\) and external fields \(h_i\). The parameter \(\beta\) is the inverse temperature.
\\
\\
In the following, we focus on the maximum a posteriori (MAP) inference problem, that is, finding the configuration
\begin{align}
x^* = \arg\max_x P(x \mid \theta).
\end{align}
In the zero-temperature limit of the Ising model (\(\beta \to \infty\)), this reduces to finding the ground state of the corresponding Ising Hamiltonian.

\subsection{Structured $M$-layer lifting}

\begin{figure*}[t]
    \centering
    \includegraphics[width=1.0\textwidth]{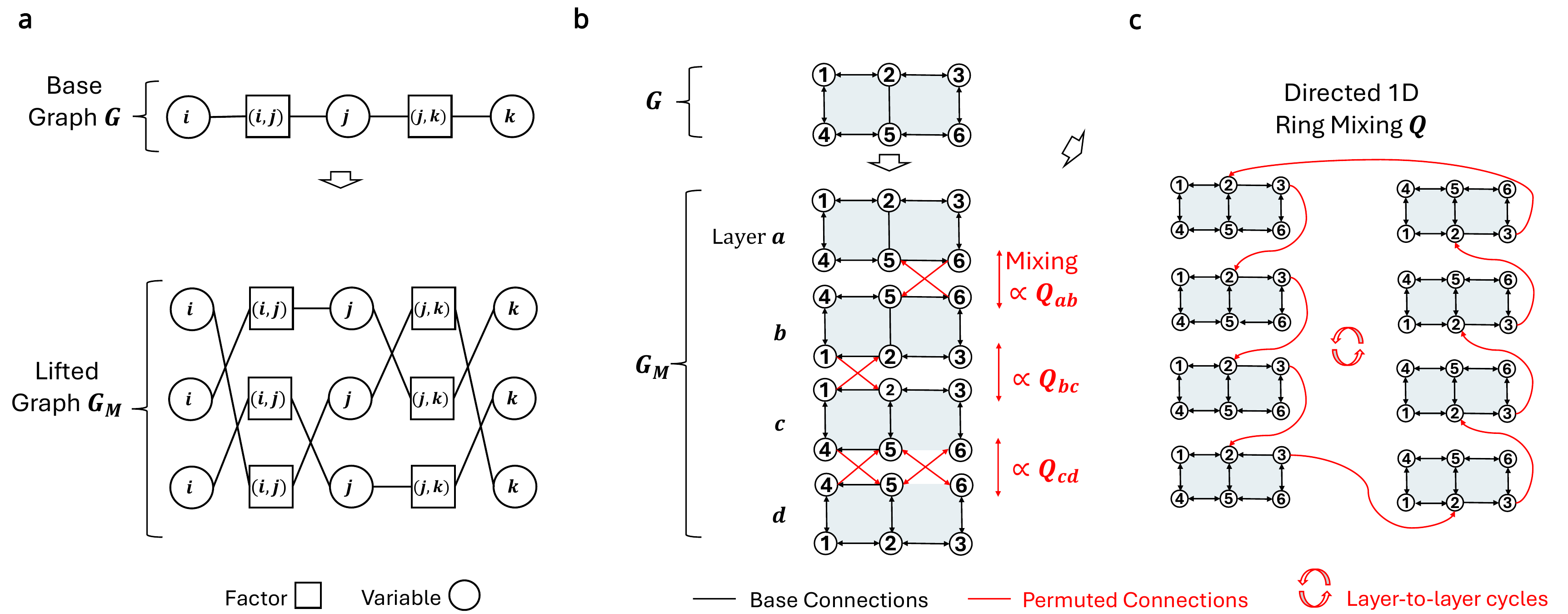}
    \caption{Schematic of the structured \( M \)-layer lift. a) $M$-layer for a 1D chain factor graph. The original graph is replicated \( M=3 \) times, with the structured graphs interconnected by directed connections permuted across layers. The quench disorder remains identical in every layer. b) Structured permutation, where the probability of permutations between blocks $a$ and $b$ is set by the mixing matrix $Q_{ab}$ in the case of an Ising model. c) $M$-layer lift obtained from a directed 1D ring $Q$ such as Gaussian circulant, illustrating that the $M$-layer lift can generate cycles in the layer-to-layer graph.}
    \label{fig:schema}
\end{figure*}

The structured $M$-layer transform extends the standard $M$-layer construction~\cite{lucibello2014finite,altieri2017loop} by introducing controlled, nonuniform mixing between layers. This additional structure allows one to shape how different copies (or layers) of the base graph are interconnected, while preserving the local interactions of the original model (see Fig.~\ref{fig:schema}).

Concretely, we introduce a mixing kernel $Q \in \mathbb{R}_{\ge 0}^{M \times M}$, where $Q_{\alpha\beta}$ sets the relative probability that a variable or factor in layer $\alpha$ connects to layer $\beta$. Uniform $Q$ recovers the standard $M$-layer construction with uniform random rewiring, while structured choices of $Q$ allow explicit control over correlations across layers.
\\
\\
The construction proceeds as follows. Starting from a base factor graph, we replicate it $M$ times, indexing each variable as $(i,\alpha)$ with $i \in \{1,\dots,N\}$ and layer index $\alpha \in \{1,\dots,M\}$. For each incidence $i \in \partial a$, we rewire connections across layers using a random permutation $\pi_{a,i}$ of $\{1,\dots,M\}$. Specifically, the factor copy at layer $\beta$ is connected to the variable copy $(i,\pi_{a,i}(\beta))$. In the asymmetric case, the two directions $i \to a$ and $a \to i$ receive independent permutations ($\pi_{a,i} \neq \pi_{i,a}$), while in the symmetric case they share the same one ($\pi_{a,i} = \pi_{i,a}$).

Structure is introduced by biasing how inter-layer permutations are drawn. Intuitively, for each directed edge
$e = (a,i)$ of the base graph $G$, the associated permutation $\pi_e$ is sampled from a matrix-weighted ensemble in which the mixing kernel $Q_{\alpha\beta}$ encodes how strongly layer $\alpha$ is biased toward connecting to layer $\beta$. In this sense, $Q$ specifies relative inter-layer preferences and can be viewed as defining an auxiliary ``graph of layers'' that governs how correlations propagate across copies. Formally, this is implemented by sampling $\pi_e$ from the distribution
\begin{align}
\label{sec:permanental}
\mathbb{P}(\pi_e) \propto \prod_{\alpha=1}^{M} Q_{\alpha,\pi_e(\alpha)},
\end{align}
which enforces a global matching between layers. Here $Q \in \mathbb{R}_{\ge 0}^{M \times M}$ is a nonnegative mixing kernel.
\\
\\
Repeating this procedure for all incidences produces a lifted graph $G_M$ with probability measure
\begin{align} \label{eq:liftedMlayer}
P_M(x) \propto \prod_{i,\alpha} f_i(x_{i,\alpha}) 
\prod_{a,\beta} f_a\Big(\{ x_{i,\pi_{a,i}(\beta)} : i \in \partial a \} \Big).
\end{align}
Comparing the lifted distribution in eq.~\eqref{eq:liftedMlayer} to the original one in eq.~\eqref{eq:Gibbs}, we see that every interaction $a$ in every layer $\beta$ in the lifted distribution exactly preserves the local neighborhood of every interaction factor $a$ in the original distribution. However a crucial difference is that the variables $i \in \partial a$ that participate in interaction $a$ can now come from different random layers in the lifted graph (see Fig.~\ref{fig:schema}).  More precisely, interaction $a$ in layer $\beta$ only involves variables $i \in \partial a$ as in the original factor graph, but the layer from which each variable $i$ originates is now random and potentially different from $\beta$, and is given by $\pi_{a,i}(\beta)$.    
For Ising models, the corresponding lifted Hamiltonian takes the form
\begin{align} \label{eq:IsingH}
H_M(x) = - \sum_{(i,j)} J_{ij} \sum_{\alpha=1}^{M} x_{i,\alpha} \, x_{j,\sigma_{ij}(\alpha)} 
- \sum_{i,\alpha} h_i \, x_{i,\alpha},
\end{align}
where the relative permutation $\sigma_{ij} = \pi_{(i,j),j} \circ \pi_{(i,j),i}^{-1}$ encodes how spins are coupled across layers (with $\sigma_{ij}=\sigma_{ji}^{-1}$ in the symmetric case).

Details on the sampling procedure for permutations are given in Appendix \ref{sec:generation}. 

\section{Empirical Results \label{sec:empirical}}

\subsection{MAP inference}

As a concrete first illustration, we apply the $M$-layer transform to an Ising model and perform zero temperature MAP inference, which is equivalent to searching for the ground-state. We start from an Ising model on a graph $G = (V, E)$ with weights $J \in \mathbb{R}^{N \times N}$, degree $d$, $N$ spins, and Hamiltonian given in eq. (\ref{eq:IsingH}). We build its structured $M$-layer lift as described in Sec.~\ref{sec:method}. We obtain the lifted graph $G_M$ with Ising coupling matrix $J_M \in \mathbb{R}^{MN \times MN}$, obtained by replicating each node $M$ times and rewiring edges according to the sampled permutations. Note that since local neighborhoods are exactly preserved in the lift, the degree of $G_M$ remains $d$. However, the total number of connections is $2M$ times that of the base graph in the asymmetric permutation case.

We measure the per-layer energy on the base Hamiltonian $H$ ($H=H_1$ using the notation of eq.~(\ref{eq:IsingH})):
\begin{align} \label{eq:edef}
e = \frac{1}{MN} \sum_{\alpha=1}^{M} H(x_{\alpha}),
\end{align}
where $\frac{e - e_0}{N}$ is the residual and $e_0$ is the ground-state energy of the base instance\footnote{The ground-state is known by construction or taken to be the best energy found across all runs when exact solve is not feasible}. Convergence time $T$ is the number of sweeps to the first local minimum.
\\
\\
The $M$-layer construction offers substantial freedom in how inter-layer permutations are chosen. Rather than analyzing this combinatorial design space directly, we focus on a simple and analytically tractable choice that already illustrates how structured lifts can affect optimization dynamics. Specifically, we adopt a simple yet nontrivial topology for the mixing matrix $Q$: a circulant ring kernel,
\begin{align}
Q_{ab} \propto \exp \left[ -\frac{(\Delta_{ab} - \mu)^2}{2\sigma^2} \right],
\end{align}
followed by row normalization, where $\Delta_{ab}$ denotes the signed circular displacement from block $a$ to block $b$, defined as $\Delta_{ab} = \bigl((b - a + B/2) \bmod B\bigr) - B/2$ for $a,b \in \{1,\ldots,B\}$. For simplifying the theoretical analysis, we will use a block-structured kernel $Q_{\alpha \beta} = Q_{b(\alpha)b(\beta)}$, where $b$ maps each layer index to one of $B$ blocks and $Q \in \mathbb{R}_{\ge 0}^{B \times B}$ is a row-stochastic matrix controlling inter- and intra-block connectivity. We typically set $M = B L$ with equal block sizes $L$, though heterogeneous partition sizes are possible.

The width $\sigma$ sets the diffusion scale, while the mean shift $\mu$ introduces a drift along the ring (see Fig.~\ref{fig:Ising_summary}a). As we show below, the ring topology induces predominantly local coupling between nearby layers, supporting long-wavelength coherent modes, while the drift $\mu$ breaks symmetry and gives rise to traveling-wave–like modes along the ring.
\\
\\
For numerical simulations, any local update rule that uses local fields/cavity messages can be run on the lifted graph such as greedy flips, Glauber updates, simulated annealing, parallel tempering and belief propagation. In the case of asymmetric permutations of the Ising model ($\pi_{a,i} \neq \pi_{i,a}$), a kinetic Ising model update rule can be utilized \cite{neri2009cavity,aurell2011message,lokhov2015dynamic}.

We start with simple Glauber algorithm for the sake of simplicity. At each Monte Carlo sweep, we calculate the local field on spin $(i,\alpha)$ and use standard single-spin updates:
\begin{align}
P(x_{i,\alpha} = +1) = \tfrac{1}{2} [1 + \tanh(\beta H_{i,\alpha})], \label{eq:glauber}
\end{align}
\noindent with
\begin{align}
H_{i,\alpha} = h_i + \sum_{j \in \partial i} J_{ij} \, x_{j,\pi(i,j)(\alpha)},
\end{align}
sweeping all $(i,\alpha)$. At $\beta = \infty$ this reduces to zero-temperature greedy flips $x_{i,\alpha} \leftarrow \operatorname{sign}(H_{i,\alpha})$. For simulated annealing, $\beta$ follows a standard schedule; for the zero-$T$ results below we run to the first local minimum (no improving flip in a full sweep).
\\
\\
In the remainder of this section, we evaluate the $M$-layer lift on three tasks: Ising ground-state search under zero-temperature dynamics, compute-to-solution of greedy dynamics and simulated annealing, and algorithmic-threshold estimates of the replica-exchange method for maximum independent set.

\subsection{Zero-temperature Quench}

We begin by performing a zero-temperature quench on the lifted graph using single-spin Glauber dynamics, and we record the average per-layer residual energy $e$, computed with respect to the base-graph Hamiltonian, upon convergence to a local minimum.

\begin{figure*}[t]
    \centering
    \includegraphics[width=1.0\textwidth]{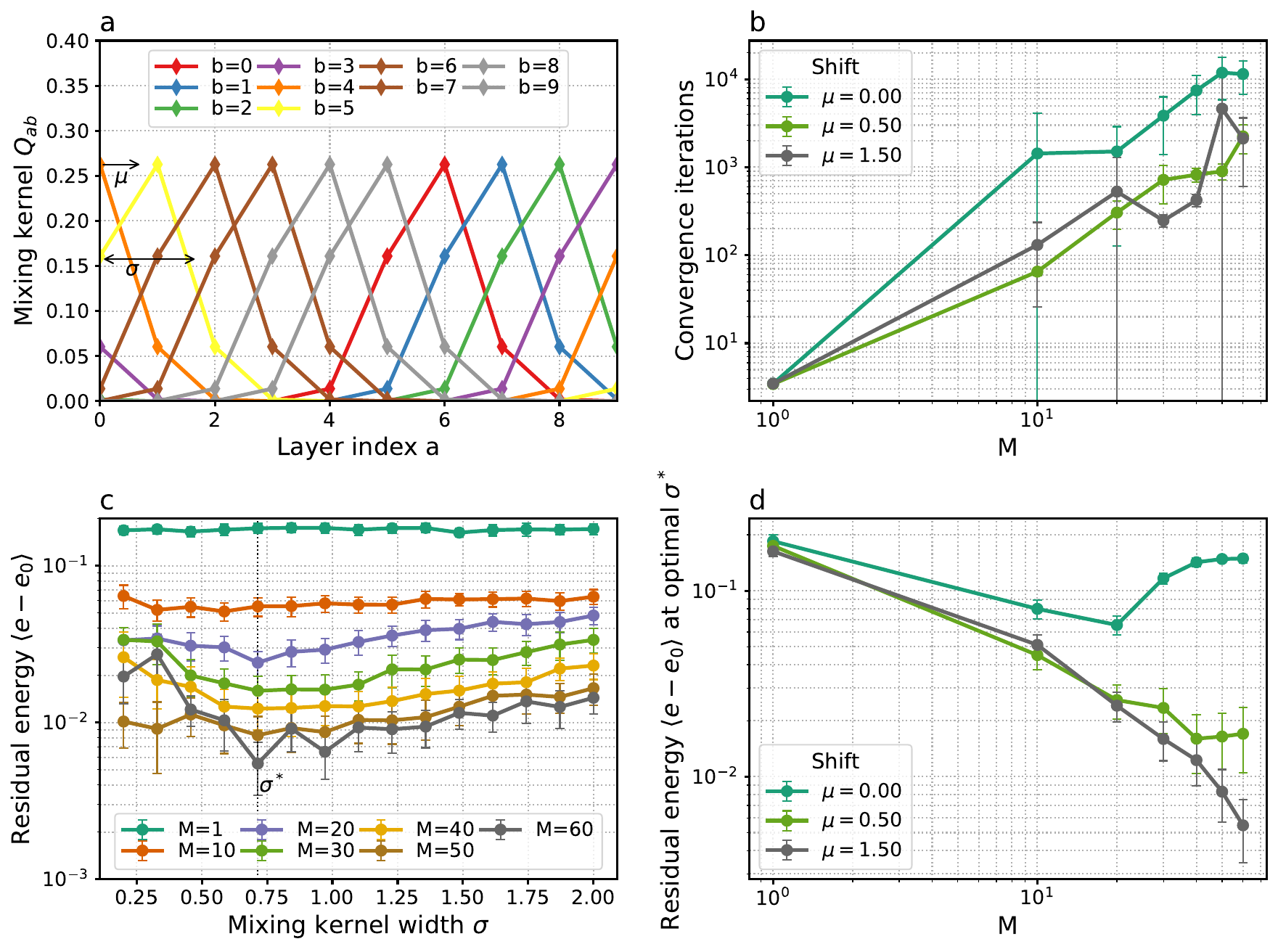}
    \caption{Zero-temperature quenches of the $M$-layer–lifted graph. a) Mixing kernel $Q_{ab}$. In this example, $Q$ is Gaussian circulant matrix with mean shift $\mu$ and width $\sigma$. b) and d) Scaling of mean convergence iterations and residual energy $\langle e - e_0 \rangle$, respectively, at optimal $\sigma^*$ with layer count $M$ for different shift values $\mu$. c) Residual energy vs. mixing kernel width $\sigma$. $L=1$. $N=50$. $T=2 \times 10^5$ maximum sweeps.}
    \label{fig:Ising_summary}
\end{figure*}

Figure~\ref{fig:Ising_summary} reports the results for lifted instances constructed from random regular base graphs of degree $d=3$ with couplings $J_{ij} \in \{-1,1\}$. Panel c) shows the average residual energy as a function of the mixing-kernel width $\sigma$ at fixed drift $\mu$. The effect of structured mixing is nonmonotonic: to minimize the residual energy, mixing must be neither too weak nor too strong. In the limit $\sigma \to 0$, the layers effectively decouple, and the dynamics reduces to independent quenched replicas of the base graph. In the opposite limit of large $\sigma$, the mixing becomes nearly uniform across layers, reproducing a fully averaged Bethe-like behavior. Between these two extremes, an optimal mixing strength $\sigma^\star$ emerges.

At this optimal $\sigma^\star$, the residual energy decreases systematically as the number of layers $M$ increases. We hypothesize that this behavior follow a power law $\langle e - e_0 \rangle \sim M^{-0.67 \pm 0.06}$ for $\mu = 1.5$ (see Fig.~\ref{fig:Ising_summary}d). 

\begin{figure*}[t]
    \centering
    \includegraphics[width=1.0\textwidth]{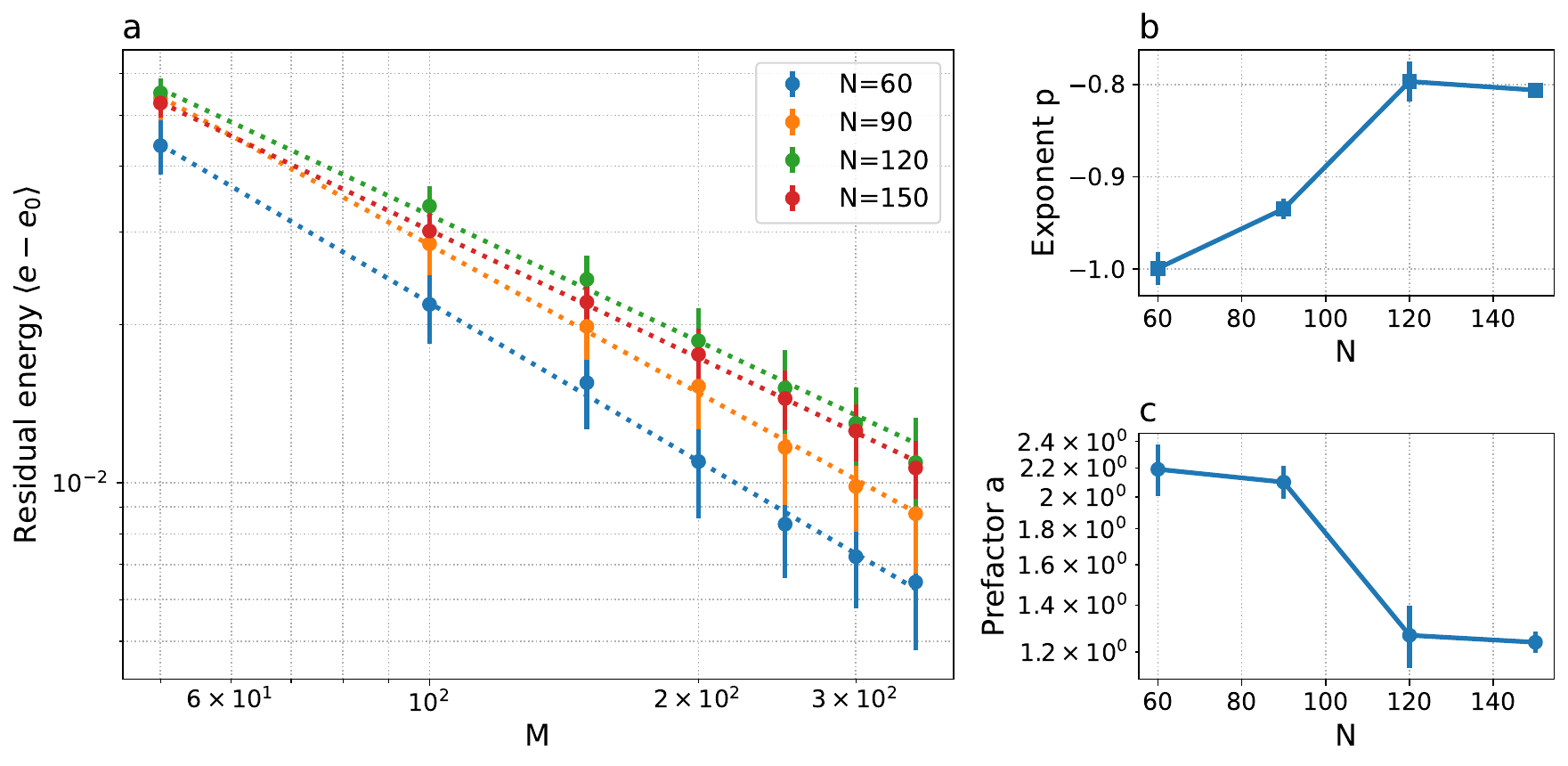}
    \caption{Decrease of residual energy with the number of layers $M$. a) Minimum over the mixing-kernel width \(\sigma\) of the mean residual energy per spin \(\langle e - e_0 \rangle\) of Ising random regular graphs of degree $k=3$, obtained by averaging \(K\) runs (and $10$ instances) of the greedy zero temperature Glauber update; vertical bars denote the stored 95\% confidence intervals, and dotted curves show the fitted power laws \(a M^p\). b) and c) Fitted exponent \(p\) versus \(N\) and prefactor \(a\) versus \(N\) on a logarithmic scale. The KS \(p\)-values are 0.157, 0.104, and 0.295 for \(N=60\), \(90\), and \(120\), respectively. $K=50$. $L=2$.}
    \label{fig:scaling}
\end{figure*}

Figure \ref{fig:scaling} shows the residual energy as a function of the number of layers $M$ for several system sizes $N$ on Bethe lattices. Empirically, a power-law fit describes the data well even for larger $N$, and Kolmogorov-Smirnov tests fail to reject the power-law model over the measured range (p-values $> 0.1$).

\subsection{Compute-To-Solution}

The structured $M$-layer lift lowers the residual energy found after greedy iterative updates but also increases the number of spins and, in turn, the number of elementary operations performed in each sweep. 
Despite this increased per-sweep cost, we find that the lift yields a net computational advantage. The probability of reaching the target energy grows sufficiently with $M$ that the total computational effort required for success is reduced. To quantify this trade-off, we measure the total computational budget, defined as the number of elementary operations needed to reach a target fraction of the ground-state energy with $99\%$ success probability, as a function of the number of blocks.

To quantify this trade-off, we define an operation-to-target (OTT) metric, which measures the total number of elementary operations required to reach the target energy with $99\%$ success probability:
\begin{align} \label{eq:budgetDef}
\text{OTT} = R \times C_{\text{run}} = R \times d\, (N M) T .
\end{align}
Here $R = \left\lceil \frac{\log(1 - 0.99)}{\log(1 - p_0)} \right\rceil$ is the number of independent runs required to achieve $99\%$ success probability, and $C_{\text{run}} = d\, (N M)\, T$ is the computational cost of a single run, proportional to the number of sweeps $T$, the number of spins $N M$, and the number of elementary operations per spin update $d$ 
($d=3$ for random regular graphs, $d=N$ for Sherrington–Kirkpatrick, and $d=4$ for tile-planted instances). The target is taken to be the inferred ground-state energy $e_0$.

For each problem type (random regular graphs, Sherrington–Kirkpatrick, and tile-planted instances \cite{perera2020chook}), we perform simulations over the key parameters: system size $N$, number of layers $M$, mixing width $\sigma$, and number of sweeps $T$. For each instance, we construct the corresponding $M$-layer couplings, run greedy and simulated annealing dynamics for $T$ sweeps, and record the per-layer energies defined in Eq.~(\ref{eq:edef}). From these runs, we estimate the empirical probability $p_0$ that at least one layer reaches the target energy level.

\begin{figure*}[t]
\centering
\includegraphics[width=1.0\textwidth]{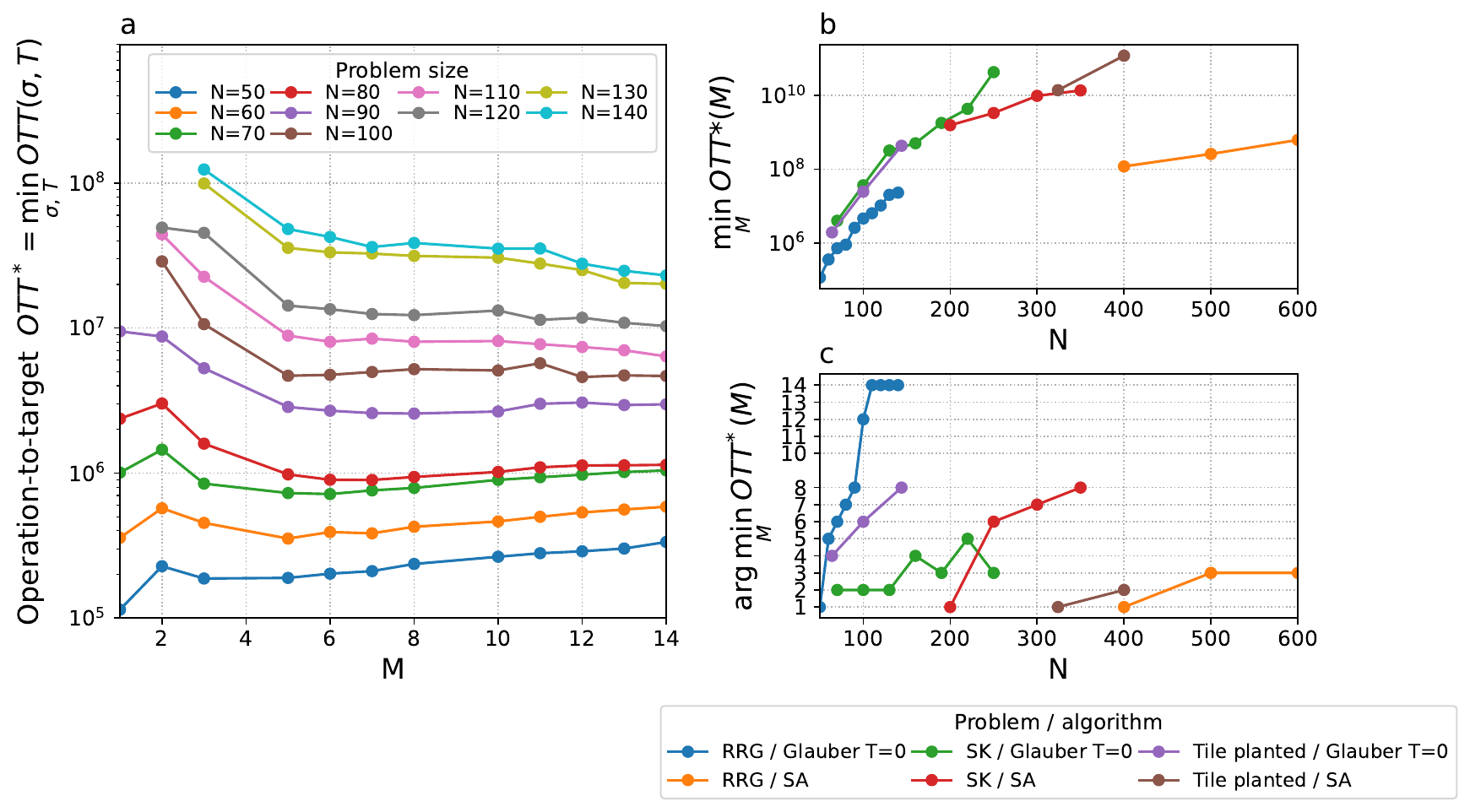}
\caption{Computational advantage of the $M$-layer lift. a) Operation-to-target (OTT) at optimal $\sigma^*$ versus layer count $M$.
b) Scaling of OTT$^*$ at optimal $M$ vs. base problem size $N$.
c) Layer count $M$ that achieves the optimal OTT for each problem and solver.
The OTT combines per-run computational cost (operations) $d N M T$ with the number of runs required for a $99\%$ success probability.
Error bars show $95\%$ confidence intervals over disorder realizations. Instances are random regular graphs of degree $d=3$. Glauber ($T=0$) = zero-temperature Glauber updates. SA = simulated annealing. RRG = random regular graph of degree $3$. SK = Sherrington-Kirkpatrick with weights $\pm 1$. Tile planted are dimension 2 with patterns $(p_1,p_2,p_3) = (0.2,0.5,0.1)$. Block size is $L=1$. The apparent saturation of the RRG/Glauber ($T=0$) curve at large $M \geq 14$ reflects the finite range of tested layer counts rather than an intrinsic limitation of the method.}
\label{fig:benchmark}
\end{figure*}

For each system size $N$, we select the minimum OTT across all tested schedules and mixing parameters, obtaining the most efficient configuration. Figure~\ref{fig:benchmark} and Table \ref{tab:OTT} summarizes these results. Across models and solvers, the OTT curves exhibit an optimum at finite $M$, confirming that structured mixing provides a computational gain beyond the trivial parallelism of repeating the system. As a general trend, the optimal number of layers $M$ increases with system size $N$, although the precise scaling depends on both the algorithm and the problem class. Further numerical studies will be required to characterize in detail how the optimal lift size depends on the underlying instance ensemble and update rule. Details of the benchmark protocol are provided in Appendix~\ref{sec:benchmark}.

\begin{table}[t]
\centering
\label{tab:ott}
\begin{tabular}{lcccccc}
\toprule
Pb. & Alg. & $\text{OTT}^*_{M=1}$ & $\min_{M} \text{OTT}^*$ & $N$ & $M^*$ & Speedup \\
\midrule
RRG & T=0 & 9.49e+06 & 2.57e+06 & 90 & 8 & 3.69 \\
RRG & SA & 1.32e+09 & 6.22e+08 & 600 & 3 & 2.13 \\
SK & T=0 & 2.22e+10 & 4.34e+09 & 220 & 5 & 5.11 \\
SK & SA & 2.47e+10 & 1.37e+10 & 350 & 8 & 1.81 \\
TP & T=0 & 4.6e+07 & 2.45e+07 & 100 & 6 & 1.88 \\
TP & SA & 1.47e+11 & 1.2e+11 & 400 & 2 & 1.22 \\
\bottomrule
\end{tabular}
\caption{\label{tab:OTT}Summary of operation-to-target (OTT) results for the $M$-layer lift. Problems, algorithm, operation-to-target at $M=1$ and optimal $M$, problem size $N$, optimal $M^*$, and speed up defined as the ratio of $\text{OTT}^*_{M=1} / \text{min}_M \text{OTT}^*(M)$. RRG: random regular graph degree $3$. SK: Sherrington-Kirkpatrick. TP: Tile-planted instances\cite{perera2020chook}.}
\end{table}


\subsection{Replica exchange methods on MIS}

To compare the structured $M$-layer lift with established replica-exchange methods, we study the maximum independent set (MIS) problem on random regular graphs. We find that the $M$-layer lift systematically raises the algorithmic threshold attainable in polynomial time when combined with replica-exchange schemes. An independent set of a graph $G=(V,E)$ is a vertex subset $S\subseteq V$ with no two vertices within $S$ connected by an edge. We define the density of a subset $S$ as $\rho = |S|/N$. The maximal independent set (MIS) is the largest possible independent set with highest density. For Monte-Carlo algorithms, a central performance measure is the number of Monte Carlo sweeps (MCS) needed to reach a given density $\rho$ of an independent set $S$ (maximal or not). 

In analyzing such algorithms, several works~\cite{angelini2019monte} introduce the notion of an algorithmic threshold $\rho_{\rm alg}$, defined as the highest independent-set density that can be reached with a runtime growing only polynomially in the system size $N$. Operationally, this threshold is identified through the scaling of the convergence time $\tau(\rho)$, defined as the typical number of iterations required to reach an independent-set density $\rho$. As $\rho$ approaches the threshold from below, the convergence time grows rapidly and is well described by a power-law divergence,
\begin{equation}
\tau(\rho)\simeq \frac{C}{(\rho_{\rm alg}-\rho)^{\nu}}.
\label{eq:alg-thresh}
\end{equation}

\begin{figure*}[t]
    \centering
    \includegraphics[width=0.80\textwidth]{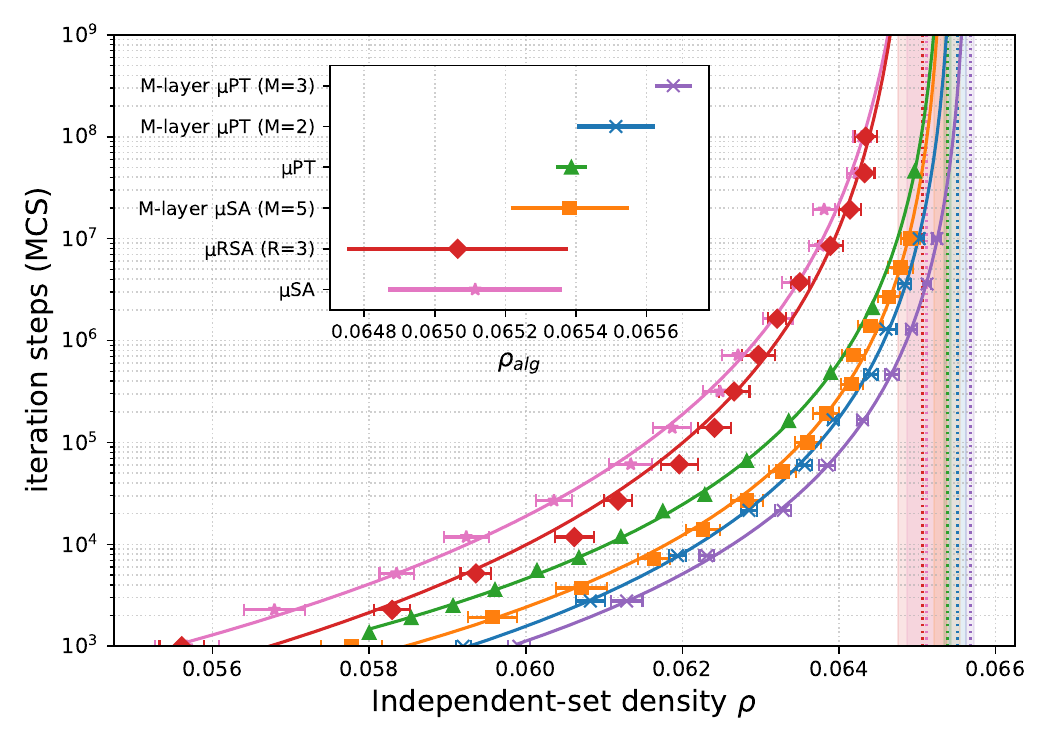}
    \caption{
    Algorithmic thresholds for MIS under replica–exchange methods.
    The main panel shows the mean number of Monte Carlo sweeps (MCS) required to reach independent-set density $\rho$ for $\mu$SA, $\mu$RSA, $\mu$PT, and their $M$-layer extensions (M-layer $\mu$SA with $M=5$, M-layer $\mu$PT with $M=3$).
    Points denote empirical measurements, and solid curves give power-law fits $\tau(\rho)=C(\rho_{\rm alg}-\rho)^{-\nu}$.
    Vertical dashed lines indicate the algorithmic thresholds $\rho_{\rm alg}$, with shaded regions showing 95\% confidence intervals.
    The inset summarizes the fitted $\rho_{\rm alg}$ values.
    The structured $M$-layer lift elevates the threshold of $\mu$SA to that of $\mu$PT and further improves $\mu$PT itself, yielding the highest threshold among all compared algorithms. $\mu$PT uses 30 replicas, M-layer $\mu$PT uses 22 and 12 replicas for $M=2$ and $M=3$, respectively.}
    \label{fig:MIS_replica_exchange}
\end{figure*}

We adopt this framework to benchmark algorithms introduced in Ref. \cite{angelini2019monte}: simulated annealing ($\mu$SA), replicated simulated annealing ($\mu$RSA), parallel tempering ($\mu$PT), and their $M$-layer extensions. Simulations are performed on random $d=100$ regular graphs of size $N=5000$ \footnote{ Although $N=5000$ is a smaller problem size than the one considered in prior works, our results are consistent with the threshold estimated in \cite{angelini2019monte} for $N=50000$.}. Figure~\ref{fig:MIS_replica_exchange} shows the measured curves and corresponding fits to eq.~\eqref{eq:alg-thresh}. For the baseline algorithms we obtain
$\mu$SA: $\rho_{\rm alg}=0.0651\pm0.0002$,
$\mu$RSA ($R=3$): $\rho_{\rm alg}=0.0651\pm0.0003$, and
$\mu$PT: $\rho_{\rm alg}=0.0654\pm0.0001$,
reproducing the established ordering in which $\mu$PT achieves the largest polynomial-time density (see Appendix \ref{sec:MISappendix} for details about the benchmark).

We then apply the structured $M$-layer lift. The lift modifies only the interaction topology and leaves each algorithm’s update rule intact. When combined with $\mu$SA using $M=5$, the threshold increases to $\rho_{\rm alg}=0.0654\pm0.0001$, effectively matching $\mu$PT. When applied to $\mu$PT with $M=3$, the threshold increases further to $\rho_{\rm alg} = 0.0657 \pm 0.0001$, the largest among all methods tested. These results show that the $M$-layer lift consistently expands the polynomial-time (``easy'') regime of both $\mu$SA and $\mu$PT, producing the strongest performance when the two approaches are combined.

\section{Cavity Theory \label{sec:mlayercavity}}

\subsection{Scope and assumption\label{sec:scope_assumption}}

Empirical results show that even simple greedy updates reach lower energies with less computational effort in a structured $M$-layer lift. To understand this gain, we analyze the lifted system within the cavity framework. These calculations extend the results of \cite{lucibello2014finite} to the structured $M$-layer setting.


Before developing the cavity-theoretic description, we clarify the assumptions under which the following analysis applies. 
With stochastic update rules, the microscopic dynamics does not admit deterministic attractors in the strict sense; rather, long-time behavior is characterized by stationary distributions and metastable regions of configuration space. In the presence of asymmetric permutations ($\pi_{a,i} \neq \pi_{i,a}$), the resulting dynamics is, in principle, non-equilibrium, and a fully general description would require dynamic cavity or message-passing equations defined on spin histories \cite{neri2009cavity,aurell2011message,lokhov2015dynamic}. Such frameworks allow for the possibility of nontrivial dynamical attractors, including oscillatory or history-dependent behavior.

In this work, however, we focus on a regime in which the structured $M$-layer mixing induces strong contraction of inter-layer fluctuations (Sec.~\ref{sec:contraction}). Empirically, for the mixing kernels $Q$ considered here, both zero-temperature and low-temperature dynamics rapidly converge to stationary configurations, and we do not observe persistent oscillatory behavior. Motivated by this observation, and by the fact that the effective coarse-grained dynamics of averaged messages becomes approximately potential-like in the large-$M$ limit, we assume that the relevant long-lived states are well described by fixed points of the belief-propagation recursion; under this assumption, the cavity and replica-symmetry-breaking analyses that follow characterize the corresponding Bethe free-energy landscape of the annealed structured $M$-layer ensemble. The resulting complexity should therefore be interpreted as the logarithmic density of Bethe fixed points, which we use as a statistical-mechanical proxy for the number of metastable basins explored by the dynamics.
\\
\\
Before turning to the technical details, we summarize the main results of this section. Under the assumptions described above, we show that the free energy of the structured $M$-layer lift reduces at leading order to a Bethe free-energy functional augmented by a linear mixing of messages across layers. The saddle-point equations of this functional yield belief-propagation (BP) updates with mixing. Linearizing these updates reveals a contraction criterion that determines when fluctuations between layers decay and the layers synchronize. Focusing on the dynamics of the averaged messages shared across layers, we further show that BP on the lifted graph can be interpreted as a stochastic descent on the Bethe free energy of the base graph. The effective noise driving this descent is controlled by coherent fluctuations across layers and enables escapes from metastable Bethe states. Finally, for the ring mixing topology, we show that introducing a drift generates traveling-wave modes across layers, which manifest as a momentum-like (Nesterov-type) acceleration of the message dynamics.

\subsection{M-layer free energy \label{sec:RSfreeenergy}}

We begin by deriving the free-energy functional of the structured $M$-layer lift. The free energy density of the structured \( M \)-layer, denoted \( f_M \) with $f_M = - \frac{1}{\beta MN} \mathbb{E}_{\pi}[\log (Z_M)]$, is computed using the replica method via:
\begin{align}
\mathbb{E}_{\pi}[\log (Z_M)] &= \lim_{n \to 0} \frac{1}{n} \left( \mathbb{E}_{\pi}[Z_M^n] - 1 \right),\\
&\approx \lim_{n \to 0} \frac{1}{n} \log (\mathbb{E}_{\pi}[Z_M^n])
\end{align}
\noindent where the expectation \(\mathbb{E}_{\pi}[\cdot]\) is taken over the random permutations \(\{ \pi_e \}\) defined at eq.~(\ref{sec:permanental}), and \(n\) is the number of replicas. The symbol $\approx$ indicates the usual replica assumption that the free energy is self-averaging, so that fluctuations of $\log Z_M$ become negligible in the thermodynamic limit.
\\
\\
We briefly outline the main steps of the calculation here; full technical details are given in Appendix~\ref{sec:structMlayer}, and a summary of notation appears in Appendix~A, Table~\ref{tab:symbols}.

We first introduce a convenient representation of the inter-layer permutations used throughout this section. For each directed edge $e=(a,i)$, the permutation $\pi_e$ defined in Eq.~(\ref{sec:permanental}) can be represented by a permutation matrix $C_e \in \{0,1\}^{M \times M}$, with entries
$C_{e,\alpha\beta}=1$ if $\alpha=\pi_e(\beta)$ and zero otherwise. 
By construction, each $C_e$ satisfies $\sum_\beta C_{e,\alpha\beta}=\sum_\alpha C_{e,\alpha\beta}=1$.
The sampling distribution of $C_e$ is then
\begin{align}
P(C_e) = \frac{1}{Z_e} \prod_{\alpha,\beta} q_{\alpha\beta}^{\,C_{e,\alpha\beta}}, 
\qquad Z_e = \mathrm{perm}(q),
\label{def:perm}
\end{align}
where $\mathrm{perm}(q)$ denotes the permanent of the mixing kernel $q$.

With this notation in place, the averaged replicated partition function is given as
\begin{align}
\mathbb{E}_C[Z_M^n]
= \sum_{\{x_a\}} \prod_{e=(i,a)} 
\sum_{\{C_e\}} P(C_e)
\prod_{\alpha,\beta} 
\left[ F_{e}^{\alpha\beta}(\{x_a\}) \right]^{C_{e,\alpha\beta}},
\end{align}
with
\begin{align}
F_{e}^{\alpha\beta} =
\exp \left[
\sum_{k=1}^n 
\left(
\log f_i(x_{i,\alpha_k})
+ 
\log f_a(x_{\partial a,\beta_k})
\right)
\right],
\end{align}
for each edge \( e = (i,a) \) and replica index \( k \).  
Here \( x_a = \{ x_i : i \in \partial a \} \) denotes the set of variables adjacent to factor \( a \).

The expression of the replicated partition function can be simplified by introducing Fourier projectors for the row/column constraints related to the permutations and performing a Hubbard--Stratonovich transform. Importantly, one can utilize the fact that only tree-like interactions in the M-layer contribute at order $M$ in $\log \mathbb{E}_{C}[Z_M^{n}]$, i.e., order $O(1)$ to the free energy density $f_M$. All subgraphs with loops contribute at higher order (see Appendix~\ref{sec:structMlayer} for details). 
\\
\\
At the end of this procedure, one obtains a functional integral over block-indexed ``messages'' (or marginal probabilities) $p$, with $p \in [0,1]^B$. From this point on, we restrict the discussion to the Ising model with pairwise factors $a=(i,j)$ for clarity. The free energy of the lifted graph depends on $\mathbb{E}_{C}[Z_M^n]$ given as
\begin{equation}
\mathbb{E}_{C}[Z_M^n]
\propto \int d\mu[p]\;\exp\!\big[-M\,S[p]\big]\ \exp[O(N)],
\label{eq:replica-functional}
\end{equation}
\noindent where the extra $\exp[O(N)]$ factor representing loop corrections is independent of $p$, contributes $O(1/M)$ to the free energy density, and can be dropped at leading order. Thus, in the limit $M \to \infty$, the free energy of the M-layer is:
\begin{align}
f_M = f_{\text{Bethe}} + O\!\left(\frac{1}{M}\right),
\end{align}
\noindent where $f_{\text{Bethe}}$ is the Bethe free energy of the M-layer with $f_{\text{Bethe}} = \frac{1}{\beta N}\,\inf_{p} S[p]$ the first order term. The action $S[p]$ is given as
\begin{equation}
S[p]
=\sum_{(i,j)\in E}\log Z_{ij}\;- \frac{1}{B} \;\sum_{i\in V} \sum_{b=1}^B \log Z_{ib},
\label{eq:action-main}
\end{equation}
with the edge and site partition factors
\begin{align}
Z_{ij}
&=\sum_{b=1}^B\sum_{x_i=\pm1}
p_b^{\,i\to j}(x_i)\,\hat p_b^{\,j\to i}(x_i),
\label{eq:Zij-main}\\
Z_{ib}
&=\sum_{x_i=\pm1}
\prod_{k\in\partial i}\hat p_b^{\,k\to i}(x_i),
\label{eq:Zi-main}
\end{align}
and propagated (block-mixed) field $\hat p_b^{\,i\to j}$ is
\begin{equation}
\hat p_b^{\,i\to j}(x_j)
=\sum_{c=1}^B Q_{bc}\sum_{x_i=\pm1}
U_{ij}(x_j,x_i)\,p_c^{\,i\to j}(x_i),
\label{eq:hatp-def-main}
\end{equation}
\noindent with $U_{ij}(\sigma,\tau)=e^{\beta J_{ij}\sigma\tau}$. The integration measure enforces normalization on each oriented edge and block as follows:
\begin{align}
d\mu[p]\;=\;
\prod_{(i,j),b}
\Bigg[\;\prod_{x=\pm1} dp_b^{\,i\to j}(x)\;
\delta\!\Big(\sum_{x} p_b^{\,i\to j}(x)-1\Big)\Bigg],
\end{align}
For $B=1$ the block index drops and eq. \eqref{eq:action-main} reduces to the standard RS Bethe free-energy functional.

\subsection{Message-passing with mixing }

We next derive the belief-propagation equations induced by the structured $M$-layer lift, which incorporate linear mixing between layers. In the thermodynamic limit \( M \to \infty \), the action \( S \) concentrates around its saddle point and yields the block-coupled belief-propagation (BP) equations (see Appendix \ref{sec:BPmixing}):
\begin{align} \label{eq:BPinp}
p_b^{\,i\to j}(x_i) &= \frac{1}{z_b^{i \to j}} \prod_{k\in\partial i\setminus j}\hat p_b^{\,k\to i}(x_i),\\
\hat p_b^{\,k\to i}&\text{ given by }\eqref{eq:hatp-def-main},
\end{align}
where $z_b^{i \to j}$ is a normalization constant. This message-passing rule generalizes standard BP by introducing block interactions through the coupling matrix \( Q \), allowing messages to mix across layers.
\\
\\
For the Ising model, it is standard to rewrite these equations in terms of scalar cavity fields. In this case, the Bernoulli message distributions belong to an exponential family, and each message can be parametrized as (see Appendix~\ref{sec:BPfield})
\begin{align}
p_b^{i \to j}(x_i) &\propto e^{\beta h_b^{i \to j} \, x_i} \label{eq:RSupdatep},\\
\hat{p}_b^{\,k\to i}(x_i) &\propto e^{\beta u_b^{\,k\to i} x_i},
\end{align}
where $h_b^{\,i \to j}$ is the cavity field sent from variable $i$ to factor $(ij)$, and $u_b^{\,k\to i}$ is the effective field propagated from neighbor $k$ to $i$.

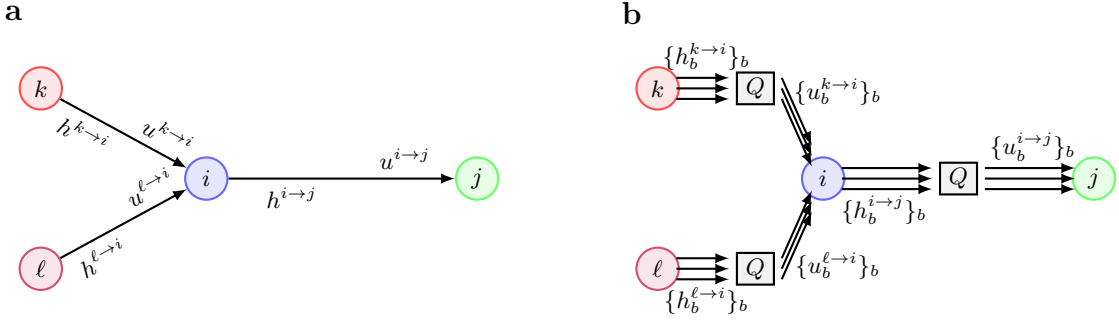
\begin{figure*}[t]
\centering
\begin{tikzpicture}[>=latex,thick,scale=1,
  var/.style  ={circle,draw,thick,minimum size=16pt,inner sep=0pt},
  block/.style={rectangle,draw,thick,minimum width=14pt,minimum height=12pt,inner sep=0pt},
  msg/.style  ={-latex,thick},
  bundle/.style={-latex,thick}
]

\node[anchor=west,font=\large] at (-2.8,2.2) {\textbf{a}};
\node[anchor=west,font=\large] at (5.4,2.2) {\textbf{b}};

\node[var,draw=blue!60,fill=blue!10]   (i)  at (0,0) {$i$};
\node[var,draw=green!60,fill=green!10] (j)  at (3.6,0) {$j$};
\node[var,draw=red!70,fill=red!10]     (k)  at (-2.2, 1.2) {$k$};
\node[var,draw=purple!70,fill=purple!10] (l) at (-2.2,-1.2) {$\ell$};

\draw[msg] (k) -- node[sloped,above,pos=0.78] {$u^{k\to i}$}
                 node[sloped,below,pos=0.26] {$h^{k\to i}$} (i);
\draw[msg] (l) -- node[sloped,above,pos=0.78] {$u^{\ell\to i}$}
                 node[sloped,below,pos=0.26] {$h^{\ell\to i}$} (i);
\draw[msg] (i) -- node[sloped,above,pos=0.78] {$u^{i\to j}$}
                node[sloped,below,pos=0.28] {$h^{i\to j}$} (j);

\begin{scope}[xshift=8.2cm]
  \node[var,draw=blue!60,fill=blue!10]     (i2)  at (0,0) {$i$};
  \node[var,draw=green!60,fill=green!10]   (j2)  at (3.6,0) {$j$};
  \node[var,draw=red!70,fill=red!10]       (k2)  at (-2.2, 1.2) {$k$};
  \node[var,draw=purple!70,fill=purple!10] (l2)  at (-2.2,-1.2) {$\ell$};

  \node[block,fill=gray!10] (Qki)  at (-0.9, 1.2) {$Q$};
  \node[block,fill=gray!10] (Qli)  at (-0.9,-1.2) {$Q$};
  \node[block,fill=gray!10] (Qij)  at ( 1.8, 0.0) {$Q$};

  \foreach \y in {-0.14,0,0.14}{
    \draw[bundle] ($(k2)+(0.25,\y)$) -- ($(Qki)+(-0.35,\y)$);
    \draw[bundle] ($(Qki)+(0.35,\y)$) -- ($(i2)+(-0.15,\y+0.3)$);
    \draw[bundle] ($(l2)+(0.25,\y)$) -- ($(Qli)+(-0.35,\y)$);
    \draw[bundle] ($(Qli)+(0.35,\y)$) -- ($(i2)+(-0.15,\y-0.3)$);
    \draw[bundle] ($(i2)+(0.25,\y)$) -- ($(Qij)+(-0.35,\y)$);
    \draw[bundle] ($(Qij)+(0.35,\y)$) -- ($(j2)+(-0.25,\y)$);
  }

  \node at ($ (k2)!0.50!(Qki) + (0, 0.42) $) {$\{h_{b}^{k\to i}\}_b$};
  \node at ($ (Qki)!0.55!(i2) + (0.6, 0.62) $) {$\{u_{b}^{k\to i}\}_b$};

  \node at ($ (l2)!0.50!(Qli) + (0,-0.42) $) {$\{h_{b}^{\ell\to i}\}_b$};
  \node at ($ (Qli)!0.55!(i2) + (0.6,-0.62) $) {$\{u_{b}^{\ell\to i}\}_b$};

  \node at ($ (i2)!0.45!(Qij) + (0,-0.42) $) {$\{h_{b}^{i\to j}\}_b$};
  \node at ($ (Qij)!0.55!(j2) + (0, 0.42) $) {$\{u_{b}^{i\to j}\}_b$};
\end{scope}

\end{tikzpicture}

\caption{Schematic of standard BP vs. structured \(M\)-layer BP with block mixing. a) Standard BP on the base graph. Each oriented edge \(e\) carries a scalar cavity field \(h^e\); its channel‑propagated counterpart is \(u^e\) (for the Ising pairwise channel \(U_{ij}(\sigma_i,\sigma_j)=e^{\beta J_{ij}\sigma_i\sigma_j}\)).  
b) Structured \(M\)-layer BP. Each oriented edge carries a block vector of cavity fields \(\{h^e_b\}_{b=1}^B\) and propagated fields \(\{u^e_b\}_{b=1}^B\). Boxes labeled \(Q\) denote the block‑mixing kernel acting linearly on the block index.
\label{fig:bp-schema}}
\end{figure*}

The node update is additive in the fields and defines a BP map $h^{+} = F(h)$, where $h^{+}$ denotes the updated cavity field after one iteration. At a fixed point it satisfies\footnote{Per-message normalization fixes the gauge, so $h_b^{\,i\to j}$ is uniquely defined.}
\begin{equation}
h_b^{\,i\to j} = F(\{h_c^{k \to i}\}_{k\in \partial i \setminus j, c})
= h_i + \sum_{k\in\partial i\setminus j} u_b^{\,k\to i}.
\label{eq:h-bp}
\end{equation}
\\
\\
At zero temperature ($T=0$), the variable--to--variable channel on $(k\to i)$ and block $b$ is
\begin{align} \label{eq:udef}
u_b^{\,k\to i}(h,J)
=\frac12\big(|J+H_b|-|J-H_b|\big),
\end{align}
\noindent and $H_b=(Qh)_b$. This corresponds to the standard zero-temperature BP field update, modified by the linear mixing induced by the matrix $Q$ \cite{mezard2003cavity} (see Fig. \ref{fig:bp-schema}).

\subsection{Bi-directional flow in the linear regime}

We next analyze the dynamics induced by the mixed belief-propagation equations by linearizing them around a fixed point. Throughout this section, we interpret BP iterations as a surrogate dynamics of the structured $M$-layer system, using them as a predictor of state stability, which we later validate with spin-level MCMC simulations. The BP dynamics is modeled as follows:
\begin{align} \label{eq:BPG}
h^+= G(h) = (1-\eta) h + \eta F(h),
\end{align}
where the left-hand side term $h^+$ is the state after one iteration and $\eta$ the update (or leak) rate introduced to improve convergence. 
\\
\\
Let $\bar{h}^*$ be a steady state of the base graph’s BP dynamics satisfying $\bar{h^*} = G(\bar{h^*})$ with $B = 1$. When $Q$ is row-stochastic, the replicated state $\bar{h} = \bar{h}^* \otimes 1_B$ is also a fixed point of the BP dynamics for any $B > 1$. We call the space $\bar{h}$ the \emph{uniform}-block manifold.
\\
\\
We analyze the stability of layer-to-layer fluctuation in the vicinity of the uniform-block manifold. On each oriented edge $e=(i\to j)$ and block $b$ we write
\begin{align} \label{eq:blockblockvar}
h_b^{i\to j}=\bar h^{\,i\to j} +\delta_b^{\,i\to j},\qquad
\sum_{b=1}^B \delta^{\,i\to j}_b=0,
\end{align}
i.e. we decompose along the all–ones direction and its zero–sum orthogonal
complement. We linearize eq.~\eqref{eq:h-bp} around \(\bar{h}\) to get
\begin{align}
\delta^+ = \left[ (1-\eta)I + \eta \left. DF \right|_{\bar h} \right] \, \delta,
\end{align}
\noindent
with
\begin{align}
\left. DF \right|_{\bar h} = (W \, K) \otimes Q,
\end{align}
where $DF$ is the Jacobian of the BP map $F$ and \( W \) is the non-backtracking operator on oriented edges of the base graph:
\begin{align}
W_{(i \to j),(k \to i)} =
\begin{cases}
1 & \text{if } k \in \partial i, \, k \ne j,\\
0 & \text{otherwise.}
\end{cases}
\end{align}
Moreover, $K = \mathrm{diag}\{\alpha_{k \to i}\}$ with \(\alpha_{k \to i}\) is the edge gain (from \(u_b^{k \to i}\) w.r.t.\ \(h_c^{k \to i}\)) evaluated at \(\bar{h}\):
\begin{align}
\alpha_{k \to i}
= \left.\frac{\partial u^{k \to i}}{\partial H^{k \to i}}\right|_{\bar{h}}
= \frac{\lambda_{ik}}{1 - \lambda_{ik}^2 \bar{m}_{k \to i}^2} \, \sech^2(\beta \bar{h}^{k \to i}),
\end{align}
with $\lambda_{ik} = \tanh(\beta J_{ik})$ and $\bar{m}_{k \to i} = \tanh(\beta \bar{h}^{k \to i})$. \(W\) encodes diffusion along the interaction graph without immediate backtracking (or non-backtracking propagation along the interaction graph) weighted by \(K\). \(Q\) mixes block components and represents propagation along the layer graph. In the linear regime, these two directions of the information flow are orthogonal.
\\
\\
\subsection{Collapse of layer-to-layer fluctuations\label{sec:contraction}}

In this subsection, we characterize the conditions under which fluctuations between layers decay and the layers synchronize to a common state. The spectral properties of $Q$ govern the dynamics of layer-to-layer fluctuations. In the general case, a sufficient criterion for synchronization ($\delta^{i \to j}_b \rightarrow 0$) is\footnote{We take $\eta=1$ for simplicity.}
\begin{align}
\rho\!\big( W K \big)\,\sigma_2(Q) < 1,
\end{align}
where $\rho(\cdot)$ denotes the spectral radius of the base-graph operator and $\sigma_2(Q)$ is the second (largest nontrivial) singular value of $Q$. 

The spectral radius $\rho(WK)$ quantifies how strongly perturbations are amplified by the dynamics on the base graph, and depends on the local couplings and fields of the current state. 
By contrast, the singular value $\sigma_2(Q)$ controls how efficiently the inter-layer mixing damps non-uniform fluctuations across layers. 
Synchronization therefore occurs when the mixing induced by $Q$ is sufficiently strong to suppress the most unstable mode of the base-graph dynamics. When this contraction condition is satisfied, the dynamics synchronizes all layers onto the block-uniform subspace in which $x_{i,\alpha}=x_i$ for all $\alpha$. 
Restricted to this subspace, the lifted Hamiltonian reduces to
$H_M(x)=M\,H(x)$, i.e., $M$ identical copies of the base-graph Hamiltonian. Minimizing $H_M$ within this synchronized subspace is therefore equivalent to minimizing $H$ on the original graph, with solutions replicated across layers.
\\
\\
More precisely, let $\{\nu_j\}$ be the eigenvalues of $W K$ (base graph interaction modes) and $\phi_k$ the ones of a normal $Q$ matrix (mixing modes). Then for every joint mode $(j,k)$, fluctuations evolve as:
\begin{align} \label{eq:decayrate}
\delta_{j,k,t+1} = \mu_{j,k} \delta_{j,k,t} \quad \text{with} \quad \mu_{jk} = (1-\eta) + \eta \nu_j \phi_k.
\end{align}
It is clear in this formulation that layer-to-layer fluctuations collapse if $\max_{j,k} |\mu_{jk}| < 1$ at the base graph state. Moreover, the decay occurs with the decay rate $|\mu_{jk}|$ with oscillation $\arg \mu_{jk}$ in each mode $(j,k)$. When \( Q \) is symmetric, \( \phi_k \in \mathbb{R} \) and each mode decays monotonically. When \( Q \) is normal but not symmetric (e.g., a circulant ring kernel with drift; see Fig. \ref{fig:Ising_summary})), its eigenmodes are complex and the BP dynamics exhibits damped oscillations of its layer-to-layer fluctuation due to wave propagation along the 1D ring (see Appendix 
\ref{sec:momentumBP}), creating a Nesterov-like momentum acceleration. This is consistent with our observation that convergence time is reduced with non-zero drift $\mu$ (see Fig. \ref{fig:Ising_summary}).
\\
\\
We emphasize that these bounds only apply under annealed mixing. If the realized mixing is a fixed symmetric permutation \(\pi\) (quenched), then
\(\sigma_2(\pi)=1\) and no extra shrinkage arises from mixing. Asymmetric permutations coupled with sequential updates at the spin-level MCMC lead to annealed mixing.

\begin{figure*}[t]
    \centering
    \includegraphics[width=1.0\textwidth]{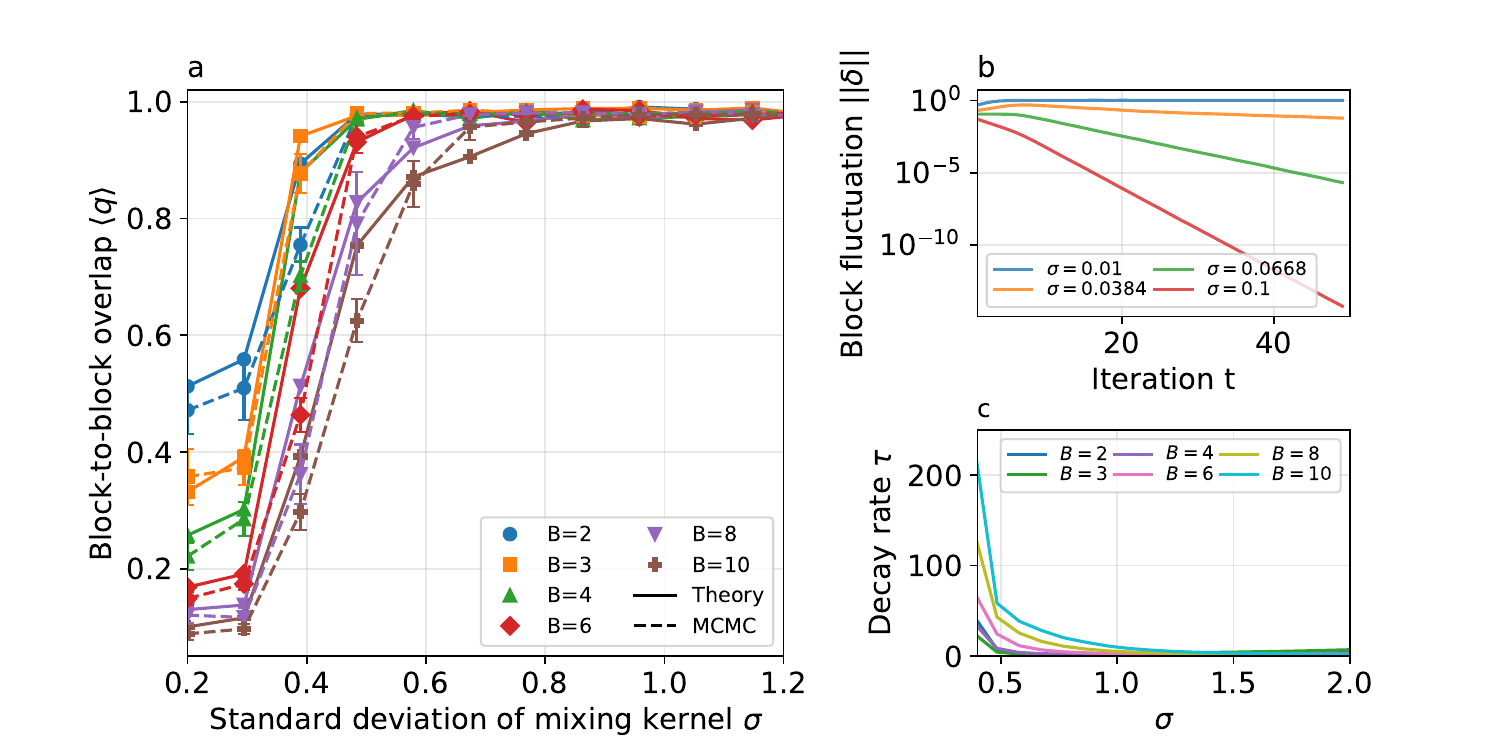}
    \caption{ Spin-level MCMC compared with cavity theory under structured mixing. a) Block-to-block overlap $\langle q \rangle$ vs.~mixing strength $\sigma$. 
    Average overlap between block from 1-RSB cavity theory and MCMC as a function of the standard deviation $\sigma$ of the mixing kernel $Q(\sigma)$. Error bars show 95\% CIs over MCMC repetitions. Theory shows the 1-RSB cavity equations at $T=0$ with Parisi parameter $y=0.31$, population $n=30000$, steps=50, and $L=30$. 
    MCMC shows $N=30$, $L=30$, with $K=30$ random initial conditions, with $200$ MCMC steps, averaged over 5 random instances. The overlap is defined as $\langle q \rangle(\sigma,B) = \langle \frac{1}{B^2} \sum_{b,b'} |q_{b,b'}| \rangle_{\rm disorder,runs} \in [0,1]$ with $q_{b,b'} = \frac{1}{N}\sum_{i=1}^{N} x_{i,b} x_{i,b'}$. Increasing $\sigma$ strengthens mixing. b) Norm of the block-to-block fluctuation $||\delta||$ (see eq. (\ref{eq:blockblockvar})) vs. iteration number of the cavity equations. c) Characteristic time $\tau$ of the block-to-block fluctuation $||\delta||$ vs. mixing strength $\sigma$ assuming $\tau = -\frac{1}{\log |\mu^*|}$ and $||\delta(t)|| = |\mu^*|^t ||\delta(0)||$ or $||\delta(t)|| = e^{-\frac{t}{\tau}} ||\delta(0)||$. $\mu^*$ corresponds to the largest eigenvalue defined in eq. (\ref{eq:decayrate}). $\eta=1$. 
    } 
    \label{fig:thresold}
\end{figure*}

\subsection{Coarse dynamics of the block average}

\begin{figure*}[t]
\centering
\includegraphics[width=\textwidth]{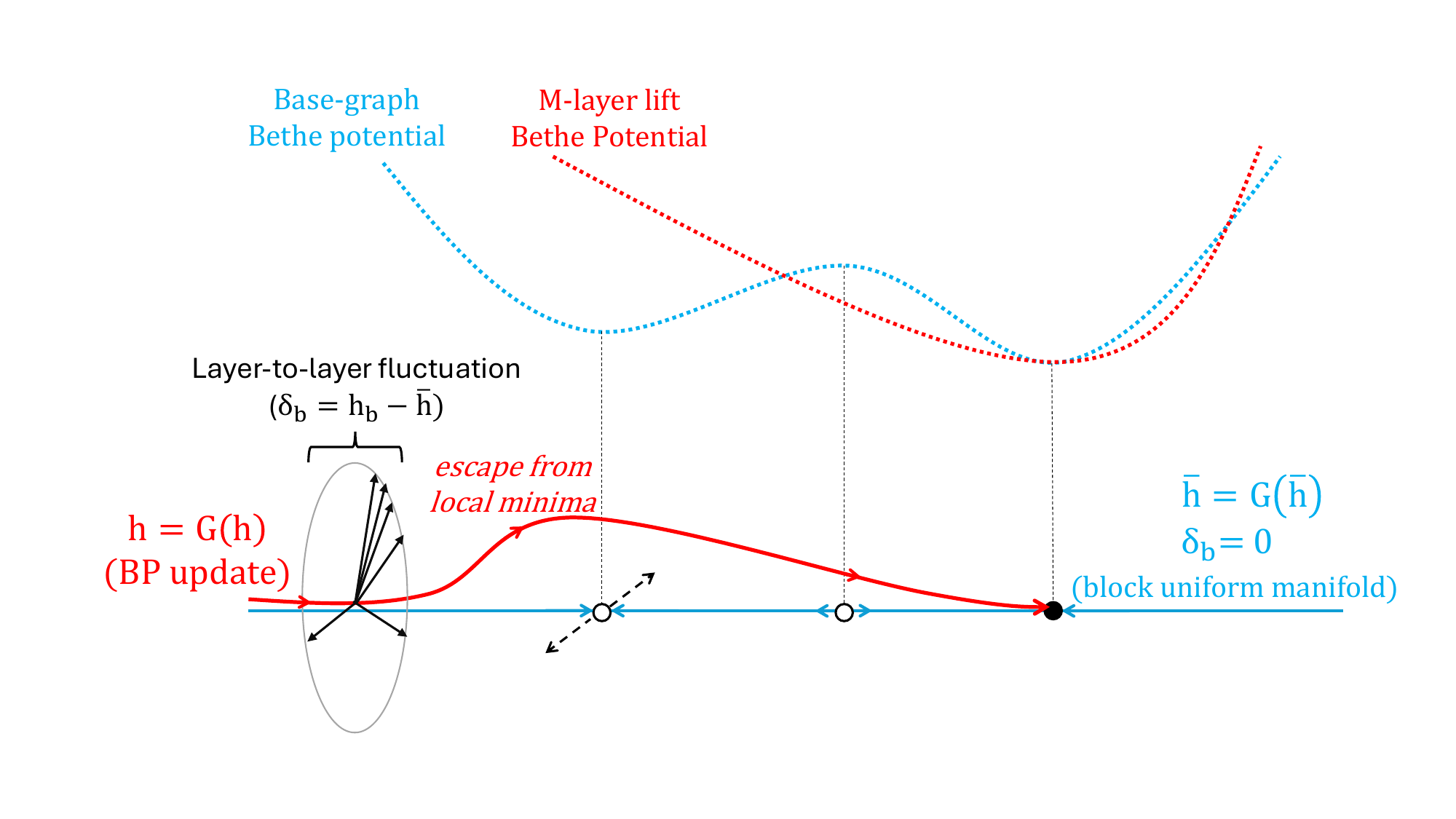}
\caption{Schematic of the dynamics on the block-uniform manifold and orthogonal layer-to-layer fluctuation subspaces.
Trajectories evolve near the uniform manifold $h_b = \bar h$ with block fluctuations $h_b = \bar h + \delta_b$. 
BP dynamics in the original graph is confined to the block uniform manifold space and could get trapped in a stable local, but not global, minimum of the base-graph Bethe potential (left minimum of the dashed blue potential and blue trajectories flowing to this minimum). However, the graph lifting leads to a higher dimensional dynamics that includes nonzero layer-to-layer fluctuations across blocks that can destabilize the previously stable BP fixed point.  This higher dimensional dynamics can thus escape local minima of the original base-graph Bethe potential (red curve), and reach to lower energies. The amplitude of the fluctuation $\delta$ is annealed near the contraction threshold.}
\label{fig:retain_sketch}
\end{figure*}

To go beyond the linear stability analysis, we examine higher-order perturbations of the belief-propagation equations and derive an effective coarse-grained dynamics for the layer-averaged messages. In Appendix \ref{sec:projhomo}, we show that the BP dynamics projected onto the block-uniform manifold (see Fig. \ref{fig:retain_sketch}) using the averaging operator  $\Pi = I \otimes \frac{1}{B}\, \mathbf{1}_B \mathbf{1}_B^{\top}$ takes the form
\begin{align}
\bar h^{+} &= G(\bar h)\;+\xi+\;O(|\delta|^3),\\
\delta^{+} &= \Pi_{\perp} DG|_{\bar h}\Pi_{\perp}\delta\;+\;O(|\delta|^2),
\end{align}
with
\begin{align}
\xi = \frac{\eta}{2B} W \big(\Gamma q_Q(\delta) - \Gamma' q_I(\delta)),
\end{align}
where 
$W$
is the non-backtracking operator on oriented edges. Moreover, $\Gamma = \operatorname{diag}(\Gamma_e)$ and $\Gamma' = \operatorname{diag}(\Gamma'_e)$ are diagonal matrices over stacked oriented edges with $\Gamma_e = \frac{2 \beta \lambda_e^3 m_e s_e^2}{ (1 - \lambda_e^2 m_e^2)^2}$ and $\Gamma'_e = \frac{2 \beta \lambda_e m_e s_e}{ 1 - \lambda_e^2 m_e^2}$ representing the edge-wise curvature (second-order sensitivity). Lastly, $q_Q(\delta) = \|Q \delta^e\|_2^2$ and $q_I(\delta) = \|\delta^e\|_2^2$ represent fluctuation power on edge $e$ with and without mixing. Although we have been able to project the dynamics of $h$ onto the dimension 1 block-uniform manifold $\bar{h}$, the noise $\xi$ still depends on higher dimensional fluctuations $\delta$.
\\
\\
The fluctuations $\delta$ are driven mainly by the residual distortion across layers of a bundle of edge messages. When the system moves between nearby states in the neighborhood of the block-uniform manifold, the part of the bundle that has not yet synchronized is reflected by a nonzero $\delta$. The noise amplitude $\xi$ thus reflects some non-local properties. The noise amplitude is driven by the difference between two terms $\Gamma q_Q(\delta)$ and $\Gamma' q_I(\delta)$. The former relates to the power of fluctuations that survive mixing while the latter is the baseline variance. Thus, $\xi$ represents the excess coherent power of layer-to-layer fluctuations after mixing.
\\
\\
When $B=1$, $\xi=0$ there are no inter-layer fluctuations. When $Q$ is homogeneous ($Q_{bc} = \frac{1}{B}$), we have $q_Q = 0$ because $Q \delta = 0$. Only the negative baseline fluctuation power remains. 
\\
\\
For structured mixing (e.g., a 1D ring), coherent fluctuations across neighboring layers yield large $q_Q(\delta)$ and therefore a strong forcing term $\xi$ facilitating escape. The noise $\xi$ acts as a layer-to-layer fluctuation coherence detector: it boosts long‑wavelengths which, in turn, triggers escape from Bethe states of the base graph. In general, cross-layer correlations that are aligned with high-gain mixer modes increase noise amplitude. 
\\
\\

\section{1-RSB Complexity \label{sec:mlayer1RSB}}

In glassy optimization problems, the energy landscape typically contains an exponential number of metastable fixed points of the replica-symmetric belief-propagation equations, often referred to as Bethe states. The associated complexity $\Sigma$ measures the logarithmic density of these states as a function of energy and provides a statistical-mechanical characterization of landscape ruggedness at the level of belief propagation.

The replica-symmetric analysis of the previous section showed that structured mixing suppresses layer-to-layer fluctuations and drives the dynamics toward block-uniform configurations. To quantify how this mechanism affects the organization and abundance of metastable Bethe states in the lifted model, we now turn to the one-step replica-symmetry-breaking (1-RSB) formalism. In this framework, complexity counts the number of distinct BP fixed points (or pure states) compatible with the mixed message-passing equations, rather than spin-level local minima of the underlying Hamiltonian.

The 1-RSB analysis thus provides a natural way to assess whether structured $M$-layer mixing merely suppresses inconsistent cross-layer configurations, or more broadly reshapes the effective ruggedness of the Bethe free-energy landscape explored by local algorithms.

\subsection{1--RSB ansatz}

For clarity, we briefly recall the one-step replica-symmetry-breaking (1-RSB) ansatz. At the RS level, each oriented edge $(i \to j)$ carries a block-indexed cavity message $p_b(x)$ parametrized by a single field $h_b$. In the 1\text{-}RSB formalism, these messages become random variables drawn from a mixture of pure states. We emphasize that the notion of layers introduced by the structured $M$-layer construction is distinct from that of replicas in the 1-RSB formalism: layers are explicitly coupled through the inter-layer permutations, whereas replicas correspond to independent copies of the lifted system that can occupy different Bethe states. Replicas are partitioned into groups of size $x$, and exchangeability within each group implies via de Finetti’s theorem that the spins in a group are i.i.d.\ conditioned on a latent cavity field. Thus the 1\text{-}RSB order parameter is a probability distribution $P_h$ over block-field vectors
\begin{align}
h = (h_1, \ldots, h_B) \in \mathbb{R}^B .
\end{align}
This distribution encodes the mixture of pure states, and the corresponding
recursive distributional equations (RDEs) arise by applying belief propagation on the auxiliary factor graph defined by the mixed BP updates.

\subsection{Auxiliary factor graph and reweighting}

Our 1-RSB analysis follows the standard cavity-theoretic framework developed in Refs.~\cite{mezard2003cavity, mezard2009information}. Interpreting RS messages as random variables allows one to define an auxiliary factor graph whose variables are $\{\, p_{i\to j},\ \hat p_{j\to i} \,\}$ and whose factors enforce the mixed BP equations $p = f(\hat p)$ and $\hat p = \hat f(p;Q)$ with $f$ and $\hat f$ defined in eqs. \eqref{eq:BPinp} and \eqref{eq:hatp-def-main}. The weight of a configuration is the Bethe free energy of the $M$-layer, raised to the Parisi exponent:
\[
W_x[p] \propto e^{x\, S[p]},
\]
where $S[p]$ is the Bethe action eq. \eqref{eq:action-main}. Applying BP on this auxiliary graph yields recursive distributional equations (RDEs) for the law $P_h$, exactly as in the standard 1-RSB cavity method but with the modification that the propagation step uses the block-mixed fields $H=Q h$.

As shown in Appendix~\ref{sec:1RSB}, regrouping the Bethe free-energy terms and choosing a convenient Bethe gauge isolates the local normalizers $z_{i\to j}$ and allows them to be written in block-factorized form. In the zero-temperature limit this produces a neighbor-separable weight, which is essential for correct RDE sampling.

\subsection{Zero-temperature RDE}

At zero temperature, we define $y=\beta x$ and take $\beta\to\infty$ with $y$ fixed and obtain the 1-RSB recursion for the distribution $P_h^{+}$ of cavity-field surveys, defined as follows:
\begin{align}
P_h^{+}(dh)
&\propto
\int 
\prod_{k \in \partial i \setminus j}
\Big[
   P_h(dh^{k})\,
   \mu_J(dJ_k)\,
   w_{\mathrm{neigh}}(h^{k}, J_k)
\Big]
\notag\\[2mm]
&\qquad\times\;
\delta\!\left(
   h - \sum_{k \in \partial i\setminus j} 
       u(h^{k}, J_k)
   \right).
\label{eq:rde-neigh-maintext}
\end{align}
with the zero-temperature neighbor weight
\begin{align}
w_{\mathrm{neigh}}(h,J)\propto
\exp\Big\{
\frac{y}{B} \sum_{b=1}^B a_b(h,J)
\Big\}.
\end{align}
The term $u_b(h,J)$ in the update rule is given in eq. (\ref{eq:udef}) and the term $a_b(h,J)$ is given as:
\begin{align}
a_b(h,J)
=\tfrac12\big(|J+H_b|+|J-H_b|\big)
=\max\{|J|,|H_b|\}.
\end{align}

For simplicity, we restrict attention to the factorized approximation of the $M$-layer cavity equations; extending the analysis to the exact 1-RSB formulation is straightforward, though it entails a substantially higher computational cost.

\subsection{Energy and complexity at $T=0$}

Moreover, the energetic potential is
\begin{equation}
\Phi(y)=
-\frac{1}{y}\log\mathbb{E}\big[e^{-y\Delta E^{(1)}}\big]
\;+\;
\frac{d}{2y}\log\mathbb{E}\big[e^{-y\Delta E^{(2)}}\big],
\end{equation}
\noindent with site- and link-energy increments given as follows:
\begin{align}
\Delta E^{(1)}
&=-\frac{1}{B}\sum_{b=1}^B\Bigg[
\sum_{k=1}^d a_b(h^{(k)},J_k)
+\Big|\sum_{k=1}^d u_b(h^{(k)},J_k)\Big|
\Bigg],\\
\Delta E^{(2)}
&= -\max_{\sigma_i,\sigma_j=\pm1}
\max_{b,b':\,Q_{bb'}>0}
\big(
h_{b,i}\sigma_i
+ J\,\sigma_i\sigma_j
+ h_{b',j}\sigma_j
\big).
\end{align}
From $\Phi(y)$ one obtains the selected energy:
\begin{align}
U^*(y)
=\frac{\partial}{\partial y}\big[y\,\Phi(y)\big],
\end{align}
\noindent and the complexity
\begin{align}
\Sigma(y)
= y\big(U^*(y)-\Phi(y)\big)
= y^2 \frac{\partial\Phi}{\partial y}.
\end{align}

\begin{figure*}[t]
    \centering
    \includegraphics[width=1.0\textwidth]{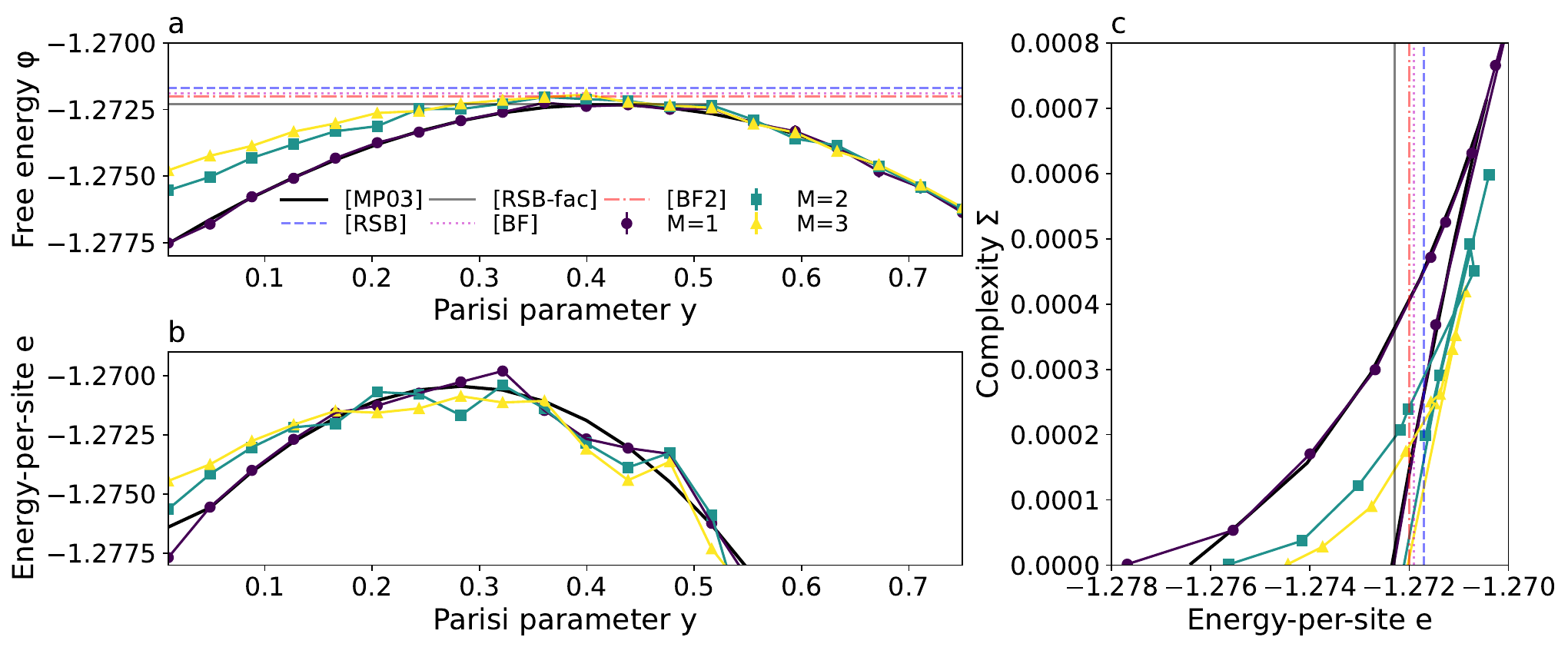}
    \caption{Complexity collapse in the structured M-layer model.
    a) Free energy $\phi(y)$ as a function of the Parisi parameter $y$. b) Energy-per-site $\epsilon(y) = \frac{d}{dy}\big(y \phi(y)\big)$. c) Complexity $\Sigma(\epsilon) = y^2 \frac{d \phi(y)}{dy}$ as a function of $\epsilon$. Black solid line [MP03] denote the factorized cavity-theory predictions from Ref.~\cite{mezard2003cavity}, while colored symbols correspond to M-layer cavity theory (this work). Reference free energy indicated are one-step replica-symmetry-breaking [RSB] with $e=-1.2717$~\cite{mezard2003cavity}, its factorized approximation [RSB-fac] with $e=-1.2723$ \cite{mezard2003cavity}, and brute force ([BF] with $e = -1.2719~$\cite{boettcher2003numerical} published in 2002, [BF2] with $e = -1.2720 $~\cite{liu2021tropical} published in 2021). Error bars represent statistical uncertainties from RDE simulations. $\mu=1.5$, $\sigma=0.8$, $L=1$. Population size is $9 \times 10^4$ cavity fields iterated for $200$ steps, averaged of 4096 trials.}
    \label{fig:complexity}
\end{figure*}

The free energy, energy-per-site, and complexity of the M-layer lift for random regular graphs of degree 3 are shown in Fig. \ref{fig:complexity}. For $B=1$ these reduce exactly to the classical 1-RSB random regular graph equations of \cite{mezard2003cavity}. As the number of blocks increase, the area under the complexity curve decreases (see Fig. \ref{fig:complexity} c)), which we interpret as a smoothing of the energy landscape induced by the interactions between layers. This collapse of complexity is consistent with the spin-level MCMC simulations in Fig. \ref{fig:scaling}, which show that increasing the number of blocks raises the probability that quench dynamics reaches the true ground state.

\section{Conclusion}

We introduced a structured $M$-layer lift that smooths the rugged energy landscape of generic probabilistic graphical models by coupling replicated factor graphs through a tunable mixing kernel. Because it modifies only the interaction topology, this construction can be combined with a broad range of local-update algorithms and problem classes, and we showed that it systematically improves access to near-MAP configurations at reduced computational cost.
\\
\\
Our cavity analysis reveals that the smoothing effect induced by the structured $M$-layer lift has a dynamical origin. When the dynamics are projected onto the block-uniform manifold of the base graph, coherent layer-to-layer fluctuations act as a finite noise source that drives controlled escapes from metastable Bethe states. Even for a minimal mixing topology such as a Gaussian circulant ring the resulting message evolution supports nontrivial behavior, including a Nesterov-like acceleration in message space.

Furthermore, the linear stability analysis identifies a contraction threshold at which the dominant inter-layer mode changes from expanding to contracting. At this point the lifetime of coherent layer fluctuations diverges, producing a transition between a disordered regime (with sustained layer heterogeneity) and a synchronized regime (in which blocks collapse onto a common trajectory). This critical behavior is reminiscent of a ferromagnetic instability, but it occurs in the higher-dimensional configuration space of cavity messages rather than in the physical spins themselves.

At the 1-RSB level, we find that the configurational complexity collapses as the number of blocks increases, providing a statistical-mechanical explanation for the observed reduction in metastable states.
\\
\\
This scheme extends the standard cavity-theoretic framework \cite{mezard2002random} by introducing a controlled linear mixing of replicated messages through a permutation ensemble parameterized by $Q$. Although one could run belief propagation directly on the lifted graph, the advantage of the structured $M$-layer lift is that it induces Bethe-like smoothing while remaining compatible with simple local dynamics such as standard MCMC. In contrast to parallel tempering or replica exchange \cite{swendsen1986replica,hukushima1996exchange}, which couple replicas by swapping thermodynamic parameters, the $M$-layer mixes the marginal information itself, producing a richer and more coherent microscopic evolution. Likewise, unlike spatially coupled LDPC constructions where a seeded boundary drives convergence, the collapse to a single solution in the structured $M$-layer arises autonomously from the inter-layer mixing.
\\
\\
While spin-level and cavity-level simulations reach near-MAP configurations on all tested instances, we do not claim polynomial-time guarantees for finding global minima in general. A more complete understanding of relaxation times, finite-size effects, and instance-dependent behavior will require a dedicated analysis. The simplicity of the construction suggests that sharper algorithmic or complexity-theoretic statements may be achievable, but such results remain open.

Our theoretical treatment focuses on the leading $O(\frac{1}{M})$ contribution to the free energy and dynamics. Higher-order terms, corresponding to explicit loop corrections in the naked-cluster expansion\cite{altieri2017loop}, have not been included here. Extending the expansion beyond first order, or designing dilution schemes that tune the loop order without relying on block structure, would provide a more complete description of finite-dilution effects. These directions lie beyond the scope of the present work but represent natural and important extensions.

We also emphasize that our theoretical analysis builds on the assumptions stated in Sec.~\ref{sec:scope_assumption}, where we restrict attention to fixed points of the mixed belief-propagation dynamics despite the presence of asymmetric interactions in the lifted graph. While this appears sufficient to capture the thermodynamically dominant behavior in the regimes studied here, a more complete treatment would require a genuinely dynamical cavity framework capable of describing time-dependent spin or message trajectories. Developing such a dynamical extension at the one-step replica-symmetry-breaking level remains an open and challenging problem.

Another promising direction is to investigate the dynamics induced by richer mixing kernels, including problem-dependent or hierarchical designs. Such kernels could shape the inter-layer correlations in more targeted ways and may further accelerate convergence toward optimal configurations.
\\
\\
Beyond spin-glass Ising models, the same lifting can likely improve combinatorial tasks such as SAT and coloring, and it may aid generalization in machine learning by favoring isotropic minima \cite{baldassi2016unreasonable}. The approach also fits hardware co-design: because the ``algorithm'' is the wiring pattern, physical implementations (e.g., optical or electronic Ising machines\cite{yamamoto2017coherent,yamamoto2020coherent}) can realize $Q$ directly as inter-layer connectivity. The same principle may also aid relaxation in other networked systems (e.g., metamaterials, power grids, social or information networks) by speeding convergence to low-energy states.






\bibliography{sample}

\subsection{Notation}

The main symbols used throughout the paper are summarized in Table~\ref{tab:symbols}.

\begin{table*}[t]
\small
\centering
\caption{Summary of main notation used in the paper. \label{tab:symbols}}
\label{tab:notation}
\begin{tabular}{l l}
\hline
Symbol & Meaning \\
\hline
$M$ & Number of layers in the lifted graph. \\
$B$ & Number of blocks used to define the structured mixing kernel. \\
$L$ & Block size, with $M = B L$. \\
$\pi_{a,i}$ & Inter-layer permutation associated with incidence $(a,i)$. \\
$Q$ & Mixing matrix controlling how edges are rewired between layers. \\
$\sigma, \mu$ & Parameters of the ring mixing kernel: width (diffusion scale) and drift (bias along the ring). \\
$H(x)$ & Ising Hamiltonian of the base graph. \\
$p_b^{\,i\to j}(x)$ & Variable-to-factor cavity message in block $b$. \\
$\hat p_b^{\,k\to i}(x)$ & Propagated (block-mixed) message from neighbor $k$ to $i$. \\
$h_b^{\,i\to j}$ & Cavity field corresponding to $p_b^{\,i\to j}$. \\
$u_b^{\,k\to i}$ & Effective field propagated from $k$ to $i$ after mixing. \\
$H_b$ & Block-mixed field defined as $H_b = \sum_c Q_{bc} h_c$. \\
\hline
\end{tabular}
\end{table*}

\subsection{Bethe functional with mixing\label{sec:structMlayer}}

In this section, we generalize the calculation of M-layer free energy\cite{lucibello2014finite} to the case of structured permutations. We work in the large-M regime $M = BL$ with fixed block count $B$ and block size $L \to \infty$, and choose the permutation weights so that individual layer-to-layer connection probabilities scale as $\mathcal{O}(1/L)$. The main objective is to average the replicated partition function of the lifted M-layer model of eq. (\ref{eq:liftedMlayer}) over the permutation ensemble  described in eq. (\ref{def:perm}) to obtain the free energy as a function of block mixing matrix $Q$ defining the matrix-weighted random permutation law.

\subsubsection{Definitions}

This section is written for the case of the Ising model (without Zeeman term) with $f_{(ij)}(x_i, x_j) = e^{\beta J_{ij} x_i x_j}$ for $x_i \in \{-1,1\}$. We thus get:
\begin{align}
P(x) = \frac{1}{Z_M} \exp(-\beta H_M(x)),
\end{align}
with $\beta$ the inverse temperature and
\begin{align}
H_M(x; C) = -\sum_{\alpha,\beta = 1}^M \sum_{i<j}^N C_{ij}^{\alpha \beta} J_{ij} x_{i \alpha} x_{j \beta}.
\end{align}

In the structured case, the probability that any two spins $i$ and $j$ in layers $\alpha$ and $\beta$, respectively, are connected, i.e., $C_{ij}^{\alpha \beta} = 1$, is parametrized by $w_{\alpha \beta}$ subject to $\sum_{\beta} C_{ij}^{\alpha \beta} = 1$ and $\sum_{\alpha} C_{ij}^{\alpha \beta} = 1$ where $C_{ij}$ is a matrix of size $M \times M$ encoding for the set of permutation between any two links $(i,j)$. We consider general $C_{ij}^{\alpha \beta}$, not necessarily symmetric. Thus, we have:
\begin{align}
P(C_{ij}) &= \frac{1}{Z_C}
\prod_{\alpha} \delta\left( \sum_{\beta} C^{\alpha\beta}_{ij} - 1 \right) 
\prod_{\beta} \delta\left( \sum_{\alpha} C^{\alpha\beta}_{ij} - 1 \right) \nonumber \\
&\times
\prod_{\alpha \beta} \Bigl[(1-w_{\alpha\beta})\,\delta(C^{\alpha\beta}_{ij})
            +w_{\alpha\beta}\,\delta(C^{\alpha\beta}_{ij}-1)\Bigr].
\end{align}

Defining the probability ratios:
\begin{equation}
q_{\alpha\beta} = \frac{w_{\alpha\beta}}{1 - w_{\alpha\beta}} \quad \text{with } 0 < w_{\alpha\beta} < 1,
\end{equation}
we have (with prefactor $\prod_{\alpha \beta} (1-w_{\alpha \beta})$ is absorbed into $Z_C$):
\begin{equation}
\begin{aligned}
P(C_{ij}) &= \frac{1}{Z_C}\,
\prod_{\alpha,\beta} q_{\alpha\beta}^{C_{ij}^{\alpha\beta}} \prod_{\alpha} \delta\!\left( \sum_{\beta} C_{ij}^{\alpha\beta} - 1 \right)
\\[6pt]
&\quad \times\,
\prod_{\beta} \delta\!\left( \sum_{\alpha} C_{ij}^{\alpha\beta} - 1 \right).
\end{aligned}
\end{equation}

This is a non-uniform doubly stochastic permutation ensemble, similar to a weighted assignment problem. $Z_C$ is a normalization constant. In the M-layer, each weight $J_{ij}$ is lifted to a matrix $J_{ij} C_{ij}$ where $C_{ij}$ is a permutation matrix of size $M \times M$. Also, note that the assignment constraints can be rewritten using the Kronecker $\delta$ as the integral representation as follows:
\begin{align}
\delta\!\Bigl(\sum_{\beta}C^{\alpha\beta}_{ij}-1\Bigr)=
   \int_{-\pi}^{\pi}\!\frac{d\lambda_{\alpha}}{2\pi}\;
   e^{\,i\lambda_{\alpha}\bigl(\sum_{\beta}C^{\alpha\beta}_{ij}-1\bigr)}.
\end{align}

\subsubsection{Averaged partition function $\mathbb{E}_{C}[Z_M^n]$}

We consider a fixed $J_{ij}$ for all layers, and calculate the average over the $M$ layer's rewiring of the replicated partition function as follows:
\begin{align}
\mathbb{E}_{C}[Z_M^{n}]
   &=
   \mathbb{E}_{C}
   \Bigl[
     \sum_{\{x^{a}\}}
     \exp\!\Bigl(
        -\beta\sum_{a=1}^{n}H_M(x^{a})
     \Bigr)
   \Bigr]\nonumber\\
   &= \sum_{\{x^a\}}
    \prod_{i<j} \sum_{\{C_{ij}\}} P(C_{ij}) \nonumber\\
   &\times \prod_{\alpha, \beta} 
    \left[
    \exp\left( \beta C_{ij}^{\alpha\beta} J_{ij} \sum_{a=1}^n x_{i\alpha}^a x_{j\beta}^a \right)
    \right].
\end{align}

Next, we plug in the probability of rewiring with the delta constraints (where $\sum_{\{C_{ij}\}}$ is for $C_{ij} \in \{0,1\}$):
\begin{align}
\mathbb{E}_{C}[Z_M^{n}]
   &= \frac{1}{Z_C^{|E|}} \sum_{\{x^a\}} \prod_{i<j} \sum_{\{C_{ij}\}} \prod_{\alpha, \beta} \left[ q_{\alpha\beta}^{C^{\alpha\beta}_{ij}} \right] \nonumber \\
    &\times
    \prod_{\alpha} \delta\left( \sum_{\beta} C^{\alpha\beta}_{ij} - 1 \right) 
    \prod_{\beta} \delta\left( \sum_{\alpha} C^{\alpha\beta}_{ij} - 1 \right) \nonumber \\
    &\times
    \prod_{\alpha, \beta} 
    \left[
    \exp\left( \beta \sum_{a=1}^n C_{ij}^{\alpha\beta} J_{ij} x_{i\alpha}^a x_{j\beta}^a \right)
    \right],\\
    &=  \frac{1}{Z_C^{|E|}} \sum_{\{x^a\}}  \prod_{i<j} \sum_{\{C_{ij}\}} \nonumber \\
    &\times
    \prod_{\alpha} \delta\left( \sum_{\beta} C^{\alpha\beta}_{ij} - 1 \right) 
    \prod_{\beta} \delta\left( \sum_{\alpha} C^{\alpha\beta}_{ij} - 1 \right) \nonumber \\
    &\times \prod_{\alpha, \beta} 
    \left[
    A_{ij}^{\alpha \beta}
    \right]^{C_{ij}^{\alpha\beta}}.
\end{align}

\noindent with $A_{ij}^{\alpha \beta} = \exp\left( \log(q_{\alpha\beta}) + \beta \sum_{a=1}^n J_{ij} x_{i\alpha}^a x_{j\beta}^a \right)$. $|E|$ is the number of edges in the original graph. We decompose the delta in Fourier series as follows:
\begin{flalign}
&\prod_{\alpha} \delta\left( \sum_{\beta} C_{ij}^{\alpha \beta} - 1 \right) \nonumber\\
&=
\int \prod_{\alpha} \frac{d\lambda_{i \to j}^{\alpha}}{2\pi} \,
\exp\left[ i \sum_{\alpha} \lambda_{i \to j}^{\alpha} \left( \sum_{\beta} C_{ij}^{\alpha \beta} - 1 \right) \right]
\end{flalign}
and get:
\begin{align}
\mathbb{E}_{C}[Z_M^{n}]
&= \frac{1}{Z_C^{|E|}} \sum_{\{x^a\}} \prod_{i<j} \int \prod_{\alpha} \left[\frac{d\lambda_{i \to j}^{\alpha}}{2\pi} \, e^{-i \lambda_{i \to j}^{\alpha}} \right] \nonumber\\
& \times \prod_{\beta} \left[\frac{d\lambda_{j \to i}^{\beta}}{2\pi} \, e^{-i \lambda_{j \to i}^{\beta}} \right] \nonumber\\
&\sum_{\{C_{ij}\}} \prod_{\alpha,\beta} \left[ e^{i \lambda_{i \to j}^{\alpha}} A_{ij}^{\alpha\beta} e^{i \lambda_{j \to i}^{\beta}} \right]^{C_{ij}^{\alpha\beta}}.
\end{align}

Moreover, we have:
\begin{flalign}
& \sum_{\{C_{ij}\}} \prod_{\alpha,\beta} \left[ e^{i \lambda_{i \to j}^{\alpha}} A_{ij}^{\alpha\beta} e^{i \lambda_{j \to i}^{\beta}} \right]^{C_{ij}^{\alpha\beta}} \nonumber\\ 
&= \sum_{\{C_{ij}\}} \prod_{\alpha,\beta} \exp\left[ C_{ij}^{\alpha\beta} \log (q_{\alpha \beta} e^{i \lambda_{i \to j}^{\alpha}} U(x_{i\alpha},x_{j\beta}) e^{i \lambda_{j \to i}^{\beta}}) \right] \nonumber\\
&= \prod_{\alpha \beta} \left( 1 + q_{\alpha \beta} e^{i \lambda_{i \to j}^{\alpha}} U_{ij}(x_{i\alpha},x_{j\beta}) e^{i \lambda_{j \to i}^{\beta}} \right) \label{eq:beforesumC}
\end{flalign}
\noindent with $U_{ij}(x_{i\alpha},x_{j\beta}) = \exp\left( \beta \sum_{a=1}^n J_{ij} x_{i\alpha}^a x_{j\beta}^a \right)$. 

\subsubsection{Linked cluster theorem and closure constraint of the M-layer lift}

For each edge $e$ of the base graph, we define $z_{e,\alpha \beta} = q_{\alpha \beta} e^{i \lambda_{i \to j}^{\alpha}} U_{ij}(x_{i\alpha},x_{j\beta}) e^{i \lambda_{j \to i}^{\beta}}$. With this notation, the replicated partition function is:
\begin{align}
\mathbb{E}_{C}[Z_M^{n}] &\propto \sum_{\{x^a\}} \int d\lambda \prod_e \prod_{\alpha \beta} (1+ z_{e, \alpha \beta})\nonumber\\
& \propto \sum_{\{x^a\}} \int d\lambda \exp \left[ \sum_{e, \alpha \beta} \log(1+ z_{e, \alpha \beta}) \right],\nonumber\\
& \propto \sum_{\{x^a\}} \int d\lambda \exp \left[ \sum_{e} \Phi_e(x,\lambda) \right],
\end{align}
with $\Phi_e = \sum_{\alpha \beta} \log(1+ z_{e, \alpha \beta})$. 

Recall that the free energy density is
\begin{align}
f_M = - \frac{1}{\beta M N} \lim_{n \rightarrow 0} \frac{1}{n} \log \mathbb{E}_{C}[Z_M^{n}].
\end{align}

We can gather the integration using $y$ with $y=(x,\lambda)$ and get:
\begin{align}
\log \mathbb{E}_{C}[Z_M^{n}] = \log \int \mathcal{D} y \exp \left( \sum_e \Phi_e(y) \right) + \text{Cst}.
\end{align}

This form is the standard one for which the linked-cluster theorem applies and gives (omitting the constant):
\begin{align}
\log \mathbb{E}_{C}[Z_M^{n}] = \sum_{\substack{H \subseteq E \\ \text{connected}}} K(H).
\end{align}

\noindent where $K (H)$ is the connected cumulant of the set $\{\Phi_e: e \in H\}$. $H$ is a connected subgraph $H$ of the base graph $G$, such as a single edge ($H=(i,j)$), or a path $H=i-j-k$, a star with 3 edges, a loop $i_1-i_2-\cdots-i_{l-1}-i_l-i1$.

Generally, the number of independent cycles of $H$, noted $l(H)$, is given as $l(H) = |E(H)|-|V(H)| + 1$. In the M-layer, we have:
\begin{align}
K(H) \sim M^{1-l(H)}.
\end{align}

To understand this, first consider a cycle of the base graph $i_1 \rightarrow i_2 \rightarrow \cdots \rightarrow i_{l-1} \rightarrow i_l \rightarrow i_1$. In the lifted M-layer graph, each edge is now dependent on a permutation $\pi_{i_k i_{k+1}} : \{1,\cdots,M\} \rightarrow \{1,\cdots,M\}$. We have in the layer index space:
\begin{align}
\alpha_1 \xrightarrow{\pi_{i_1 i_2}} \cdots \xrightarrow{\pi_{i_{l-1} i_l}} \alpha_l \xrightarrow{\pi_{i_{l} i_1}} \alpha_1'.
\end{align}

To close the loop we need $\alpha_1 = \alpha_1'$, that is we need the permutation $\alpha_1' = (\pi_{i_l i_1} \circ \cdots \pi_{i_1 i_2}) (\alpha_1)$ to map back to itself. The probability that the lifted layer index is consistent around one base-graph cycle is $\frac{1}{M}$. The probability of closure of $l$ loops is $\frac{1}{M^l}$ for $l \ll M$ where $l$ is the number of independent cycles.

Then, start from any root vertex $i_1$. There is $M$ possible choices of layers. The total contribution of a subgraph of the base graph with $l$ loop is thus $M \times \frac{1}{M^l}$ in the lifted graph.

For example, trees ($l(H)=0$) contribute at order $M$ in $\log \mathbb{E}_{C}[Z_M^{n}]$, i.e., order $O(1)$ to the free energy density $f_M$ (this is the Bethe free energy). One cycle contributes $O(\frac{1}{M})$, etc.

We thus have 
\begin{align} \label{eq:treesubgraphs}
\log \mathbb{E}_{C}[Z_M^{n}] = \sum_{\substack{T \subseteq E \\ \text{connected}}} K(T) + O(N)
\end{align}
\noindent with $T$ the set of tree subgraphs with $l(T)=0$ of the base graph $E$. In the following, we will denote by $f_{\text{Bethe}}$ the first order in the free energy such that:

\begin{align}
f_M = f_{\text{Bethe}} + O(\frac{1}{M}),
\end{align}

\noindent where $f_{\text{Bethe}}$ is formally defined as $f_{\text{Bethe}} = - \frac{1}{\beta M N} \lim_{n \rightarrow 0} \frac{1}{n} (\sum_{\substack{T \subseteq E \\ \text{connected}}} K(T)$). 

\subsubsection{Bethe functional at first order}

In the large-lift regime with per-entry scaling $z_{e,\alpha\beta} = O(1/L)$ since $q_{\alpha\beta} = O(1/L)$, one has $\log(1 + z_{e,\alpha\beta}) = z_{e,\alpha\beta} + O(L^{-2})$, so higher powers in the local expansion are subleading. Moreover, the linked-cluster expansion organizes $\log \mathbb{E}_{C}[Z_{M}^{n}]$ as a sum over connected base subgraphs $H$, whose contributions scale as $K(H) \sim M^{1-l(H)}$; thus loop subgraphs ($l(H) \ge 1$) yield only $O(1/M)$ corrections to the free-energy density. Thus, we thus get at the first order by replacing $\log (1+z_{e,\alpha \beta})$ by $z_{e,\alpha \beta}$:
\begin{align}
\mathbb{E}_{C}[Z_M^{n}]
&= \frac{1}{Z_C^{|E|}} \sum_{\{x^a\}} \prod_{i<j} \int \prod_{\alpha} \left[\frac{d\lambda_{i \to j}^{\alpha}}{2\pi} \, e^{-i \lambda_{i \to j}^{\alpha}} \right] \nonumber \\
& \times \prod_{\beta} \left[\frac{d\lambda_{j \to i}^{\beta}}{2\pi} \, e^{-i \lambda_{j \to i}^{\beta}} \right] \nonumber \\
&\times \exp \left[\sum_{\alpha \beta} q_{\alpha \beta} e^{i \lambda_{i \to j}^{\alpha}} U_{ij}(x_{i\alpha},x_{j\beta}) e^{i \lambda_{j \to i}^{\beta}} \right] \nonumber \\
& \times (1+O(\frac{1}{M}))^{MN},
\end{align}
\noindent where the left hand side $(1+O(\frac{1}{M}))^{MN}$ reflects the fact that we consider only the tree clusters at first order.

For comparison, in the standard $M$-layer construction with uniform random permutations\cite{lucibello2014finite}, we have $w_{\alpha \beta} = \frac{1}{M}$, i.e., $q=\frac{1}{M-1}$, and probability of $C_{ij}^{\alpha\beta} = 1$ ($C_{ij}^{\alpha\beta}=0$) is $w_{\alpha \beta}$ ($1-w_{\alpha \beta}$):

\begin{align}
\mathbb{E}_{C}[Z_M^{n}]
&= \frac{1}{Z_C^{|E|}} \sum_{\{x^a\}} \prod_{i<j} \int \prod_{\alpha} \left[\frac{d\lambda_{i \to j}^{\alpha}}{2\pi} \, e^{-i \lambda_{i \to j}^{\alpha}} \right] \nonumber \\
& \times \prod_{\beta} \left[\frac{d\lambda_{j \to i}^{\beta}}{2\pi} \, e^{-i \lambda_{j \to i}^{\beta}} \right] \nonumber \\
&\times \exp \left( q \sum_{\alpha \beta} e^{i \lambda_{i \to j}^{\alpha}} U_{ij}(x_{i\alpha},x_{j\beta}) e^{i \lambda_{j \to i}^{\beta}} \right) \nonumber \\
&\times(1+O(\frac{1}{M}))^{MN}.
\end{align}

From there, calculation to show that the free energy of the M-layer converges to the Bethe free energy can be continued as in \cite{lucibello2014finite}.

\subsubsection{Structured permutation}

Let the $M$ layers be partitioned into $B$ equally-sized blocks of size $L$ (so $M = BL$). We introduce a block–membership map
\[
b : \{1, \dots, M\} \to \{1, \dots, B\}, \quad b(\alpha) = 1 + \left\lfloor \frac{\alpha - 1}{L} \right\rfloor,
\]

\noindent or any other fixed rule that assigns every layer index $\alpha$ to a block label $b(\alpha)$.

An arbitrary non-negative kernel \( Q_{b b'} \) is chosen with row-stochastic (\( \sum_{b'} Q_{b b'} = 1 \)) and with spectral radius \( \leq 1 \). Define the structured weights:

\begin{align}
q_{\alpha\beta}\;=\;Q_{\,b(\alpha)b(\beta)}.
\end{align}

First, we insert this ansatz into eq.~(\ref{eq:beforesumC}). After introducing the row/column Kronecker constraints via their Fourier representations, the sum over $C_{ij}$ factorizes and we obtain, at first order (Bethe approximation, see previous discussion of the linked-cluster theorem), the replacement:
\begin{align}
\prod_{\alpha,\beta}\Bigl(1+ q_{\alpha\beta}\,X^{\alpha\beta}_{ij}\Bigr)
&\gets
\exp\!\Bigl[\sum_{\alpha,\beta} q_{\alpha\beta}\,X^{\alpha\beta}_{ij}\Bigr] \nonumber\\
& =
\exp\!\Bigl[\sum_{b,b'} Q_{bb'}\,X^{bb'}_{ij}\Bigr],
\end{align}
where
\begin{align}
X^{\alpha\beta}_{ij} &= e^{i\lambda^{\,i\to j}_{\alpha}}\,
U_{ij}\bigl(x_{i\alpha},x_{j\beta}\bigr)\,
e^{i\lambda^{\,j\to i}_{\beta}}, \\
X^{bb'}_{ij} &=\sum_{\alpha\in b}\sum_{\beta\in b'} X^{\alpha\beta}_{ij},
\end{align}
and
\begin{align}
U_{ij}\bigl(x_{i\alpha},x_{j\beta}\bigr)
=\exp\!\Bigl(\beta J_{ij}\sum_{a=1}^n x_{i\alpha}^ax_{j\beta}^a\Bigr).
\end{align}

Therefore,
\begin{align}
\mathbb{E}_{C}[Z_M^{n}]
&= \frac{1}{Z_C^{|E|}}\,
\sum_{\{x^a\}}
\prod_{i<j}
\int\!
\Bigg[\prod_{\alpha}\frac{d\lambda_{i\to j}^{\alpha}}{2\pi}\,e^{-i\lambda_{i\to j}^{\alpha}}\Bigg] \nonumber\\
& \times \Bigg[\prod_{\beta}\frac{d\lambda_{j\to i}^{\beta}}{2\pi}\,e^{-i\lambda_{j\to i}^{\beta}}\Bigg]
\nonumber\\
&\times\,
\exp\!\Bigl[\sum_{b,b'} Q_{bb'}\,X^{bb'}_{ij}\Bigr] (1+O(\frac{1}{M}))^{MN}.
\end{align}
\noindent

We can also simply write the last term as $\exp \Bigl[\sum_{\alpha,\beta} q_{\alpha \beta}\,X^{\alpha \beta}_{ij}\Bigr]$ for any mixing matrix $q$ (not necessarily block matrix).


\subsubsection{Empirical-field representation (block form)}

For each oriented edge $(i\to j)$ and block $b$, define the empirical fields over replica spin vectors
$x, \tau \in \{\pm1\}^n$:
\begin{align}
\hat{\rho}^{\,i \to j}_b(x) &\;=\;
\sum_{\alpha:\, b(\alpha)=b} e^{i\lambda^{\,i \to j}_\alpha}\,
\delta_{x,\;x^i_\alpha},\\
\check{\rho}^{\,j \to i}_b(\tau) &\;=\;
\sum_{\beta:\, b(\beta)=b} e^{i\lambda^{\,j \to i}_\beta}\,
\delta_{\tau,\;x^j_\beta},
\end{align}
where $\delta_{\cdot,\cdot}$ is the Kronecker delta on $\{\pm1\}^n$.

For every edge $(i,j)$,
\begin{align}
\sum_{\alpha,\beta} X^{\alpha\beta}_{ij}
&=\sum_{\alpha,\beta} e^{i\lambda^{\,i \to j}_\alpha}\,
U_{ij}\!\big(x^i_\alpha,x^j_\beta\big)\,
e^{i\lambda^{\,j \to i}_\beta}\nonumber\\
&=\sum_{b,b'} \sum_{x,\tau\in\{\pm1\}^n}
\hat{\rho}^{\,i \to j}_b(x)\,
U_{ij}(x,\tau)\,
\check{\rho}^{\,j \to i}_{b'}(\tau) \nonumber\\
&= \sum_{b,b'} \big\langle
\hat{\rho}^{\,i \to j}_b,\;U_{ij},\;\check{\rho}^{\,j \to i}_{b'}\big\rangle,
\end{align}
with 
\begin{align}
U_{ij}(x,\tau) &=\exp\!\Big(\beta J_{ij}\sum_{a=1}^n x^{a}\tau^{a}\Big),\\
\langle \hat{\rho},U,\check{\rho}\rangle
&=\sum_{x,\tau}\hat{\rho}(x)\,U(x,\tau)\,\check{\rho}(\tau).
\end{align}

Using the block-structured couplings $q_{\alpha\beta}=Q_{b(\alpha)b(\beta)}$, the replicated average becomes
\begin{align}
\mathbb{E}_{C}[Z_M^{n}]
&=\frac{1}{Z_C^{|E|}}
\sum_{\{x^a\}}
\prod_{i<j}
\int\!\Bigg[\prod_{\alpha}\frac{d\lambda^{\,i \to j}_\alpha}{2\pi}\,e^{-i\lambda^{\,i \to j}_\alpha}\Bigg] \nonumber \\
\times & \Bigg[\prod_{\beta}\frac{d\lambda^{\,j \to i}_\beta}{2\pi}\,e^{-i\lambda^{\,j \to i}_\beta}\Bigg]
\nonumber\\
&\times
\exp\!\left[\sum_{b,b'} Q_{bb'}\,
\big\langle \hat{\rho}^{\,i \to j}_b,\;U_{ij},\;\check{\rho}^{\,j \to i}_{b'}\big\rangle\right] \nonumber\\
&\times (1+O(\frac{1}{M}))^{MN}.
\label{eq:rhoeq}
\end{align}

\subsubsection{Decoupling the quadratic term}

We now decouple the bilinear form in eq. (\ref{eq:rhoeq}) by means of a Hubbard–Stratonovich transform.

Let
\begin{align}
K_{ij}^{\,bb'}(x,\tau) \;=\; Q_{bb'}\,U_{ij}(x,\tau),
\end{align}
and recall that
\begin{align}
U_{ij}(x,\tau) \;=\; \exp\!\left(\beta J_{ij}\sum_{a=1}^n x^a\tau^a\right),
\quad x,\tau\in\{\pm1\}^n.
\end{align}
We denote the quadratic form
\begin{align}
\langle \hat\rho,\,K_{ij},\,\check\rho\rangle
\;=\;
\sum_{b,b'=1}^B \;\sum_{x,\tau}
\hat\rho_b(x)\,K_{ij}^{\,bb'}(x,\tau)\,\check\rho_{b'}(\tau),
\end{align}
where the sums over $x,\tau$ are over the discrete replica spin space $\{\pm1\}^n$.

For any invertible $K$ acting on the $(b,x)$ space\footnote{When $Q$ is symmetric positive definite, the HS integral is convergent in the usual sense.},
\begin{align}
\exp\big[\langle \hat\rho,\,K,\,\check\rho\rangle\big] &\propto
\int\!\mathcal{D}\varphi\,\mathcal{D}\psi\;
\exp [
 -\langle \varphi,\,K^{-1},\,\varphi\rangle \nonumber\\
 &-\langle \psi,\,K^{-1},\,\psi\rangle 
 +\langle \hat\rho,\,\varphi+i\psi\rangle \nonumber\\
 &+\langle \check\rho,\,\varphi-i\psi\rangle
],
\label{eq:HS}
\end{align}
where $\varphi,\psi$ are real fields $\varphi_b(x)$, $\psi_b(x)$,  
and the Gaussian measures $\mathcal{D}\varphi$, $\mathcal{D}\psi$ integrate over $\mathbb{R}^{B\times 2^n}$.

We apply \eqref{eq:HS} with
$\hat\rho_b=\hat\rho^{\,i\to j}_b$ and $\check\rho_{b'}=\check\rho^{\,j\to i}_{b'}$.
This yields, for each oriented edge $(i\to j)$,
\begin{widetext}
\begin{align}
&\exp\left[
\sum_{b,b'} Q_{bb'}\,
\langle \hat\rho^{\,i\to j}_b,\,U_{ij},\,\check\rho^{\,j\to i}_{b'}\rangle
\right] \propto
\int\!\mathcal{D}\varphi^{\,i\to j}\,\mathcal{D}\psi^{\,i\to j}\;
\exp\Big[
 -\sum_{b,b'}\langle \varphi^{\,i\to j}_b,\,Q^{-1}_{bb'}\otimes U_{ij}^{-1},\,\varphi^{\,j\to i}_{b'}\rangle \\
&\hspace{37mm}
 -\sum_{b,b'}\langle \psi^{\,i\to j}_b,\,Q^{-1}_{bb'}\otimes U_{ij}^{-1},\,\psi^{\,j\to i}_{b'}\rangle
 +\sum_b\sum_x \hat\rho^{\,i\to j}_b(x)\,[\varphi^{\,i\to j}_b(x)+i\psi^{\,i\to j}_b(x)] \nonumber\\
&\hspace{37mm}
 +\sum_{b'}\sum_\tau \check\rho^{\,j\to i}_{b'}(\tau)\,[\varphi^{\,j\to i}_{b'}(\tau)-i\psi^{\,j\to i}_{b'}(\tau)]
\Big].\nonumber
\end{align}
\end{widetext}

Next, we define the complex fields
\begin{align}
\rho^{\,i\to j}_b(x) &= \varphi^{\,i\to j}_b(x) + i\,\psi^{\,i\to j}_b(x), \\
\rho^{\dagger\,j\to i}_b(\tau) &= \varphi^{\,j\to i}_b(\tau) - i\,\psi^{\,j\to i}_b(\tau).
\end{align}
Then
\begin{align}
\sum_b\sum_x \hat\rho^{\,i\to j}_b(x)\,\rho^{\,i\to j}_b(x)
=\sum_{\alpha=1}^M e^{i\lambda^{\,i\to j}_\alpha}\,
\rho^{\,i\to j}_{\,b(\alpha)}(x_{i\alpha}),
\end{align}
and similarly for the column term with $\rho^\dagger$.

The $\lambda$–integrals factorize over $\alpha$:
\begin{align}
\int_{-\pi}^{\pi}\!\frac{d\lambda}{2\pi}\,e^{-i\lambda} e^{z e^{i\lambda}} = z,
\end{align}
so that
\begin{flalign}
&\int\!\prod_{\alpha}\frac{d\lambda^{\,i\to j}_\alpha}{2\pi} e^{-i\lambda^{\,i\to j}_\alpha}
\exp\!\left[\sum_{\alpha} e^{i\lambda^{\,i\to j}_\alpha}\rho^{\,i\to j}_{\,b(\alpha)}(x_{i\alpha})\right] \nonumber\\
&=\prod_{\alpha=1}^M \rho^{\,i\to j}_{\,b(\alpha)}(x_{i\alpha}),
\end{flalign}
and likewise for the $(j\to i)$ factors with $\rho^\dagger$.

For each $(i,j)$, we have
\begin{flalign}
& \int\!\mathcal{D}\rho^{\,i\to j}\,\mathcal{D}\rho^{\dagger\,j\to i}\;
\exp\!\left[ -\sum_{b,b'}\langle \rho^{\,i\to j}_b,\,Q^{-1}_{bb'}\otimes U_{ij}^{-1},\,\rho^{\dagger\,j\to i}_{b'}\rangle \right] \nonumber\\
&\times\;
\prod_{\alpha=1}^M \rho^{\,i\to j}_{\,b(\alpha)}(x_{i\alpha})
\;\prod_{\beta=1}^M \rho^{\dagger\,j\to i}_{\,b(\beta)}(x_{j\beta}).
\end{flalign}

\subsubsection{Change of variables}

We perform a change of variable by introducing new fields $p^{\,j\to i}_c(\tau)$ via
\begin{equation}
\rho^{\,i\to j}_b(x) \;=\;
\sqrt{M}\sum_{c=1}^B Q_{bc}\sum_{\tau} U_{ij}(x,\tau)\,p^{\,j\to i}_c(\tau),
\label{eq:rho-to-p}
\end{equation}
and similarly for $\rho^\dagger$ with $p^\dagger$.
This transformation is linear with Jacobian
\begin{align}
\mathcal{D}\rho\,\mathcal{D}\rho^\dagger \propto
\mathcal{D}p\,\mathcal{D}p^\dagger.
\end{align}
The quadratic form transforms as
\begin{align}
-\sum_{b,b'}\langle \rho_b,\,Q^{-1}_{bb'}\otimes U_{ij}^{-1},\,\rho^\dagger_{b'}\rangle
= -M\,\sum_{b,b'} Q_{bb'}\,\langle p_b^{\,i\to j},\,U_{ij},\,p_{b'}^{\dagger\,j\to i}\rangle,
\end{align}
with
\begin{align}
\langle p_b,\,U_{ij},\,p_{b'}^\dagger\rangle
=\sum_{x,\tau} p_b(x)\,U_{ij}(x,\tau)\,p_{b'}^\dagger(\tau).
\end{align}

Putting everything together, the replicated, $C$–averaged partition function reads
\begin{align}
\mathbb{E}_{C}[Z_M^{n}]
&\propto 
\sum_{\{x^a\}}
\int\!\prod_{(i,j)}\prod_{b=1}^B \mathcal{D}p_b^{\,i\to j}\,\mathcal{D}p_b^{\dagger\,j\to i} \nonumber\\
&\quad\times
\exp\!\left[
- M \sum_{(i,j)}\sum_{b,b'} Q_{bb'}\,
\langle p_b^{\,i\to j},\,U_{ij},\,p_{b'}^{\dagger\,j\to i}\rangle
\right] \nonumber\\
&\quad\times
\prod_{(i,j)}
[
\prod_{\alpha=1}^M \left(U_{ij}Q\,p^{\,j\to i}\right)_{b(\alpha)}(x_{i\alpha}) \nonumber\\
&\quad\times \prod_{\beta=1}^M \left(U_{ij}Q\,p^{\,i\to j}\right)_{b(\beta)}(x_{j\beta})
] (1+O(\frac{1}{M}))^{MN},
\end{align}
\noindent with
\begin{align}
\left(U_{ij}Q\,p^{\,j\to i}\right)_b(x)
=\sum_{c=1}^B Q_{bc}\sum_{\tau}U_{ij}(x,\tau)\,p_c^{\,j\to i}(\tau).
\end{align}

\subsubsection{Gather over configurations}

For each oriented edge $(j \to i)$ and block index $b\in\{1,\dots,B\}$, define
\begin{align}
F_b^{\,j \leftarrow i}(x)
&= \left(U_{ij} Q\, p^{\dagger\,j \leftarrow i}\right)_b(x) \nonumber \\
&= \sum_{c=1}^B Q_{bc} \sum_{\tau} U_{ij}(x,\tau)\,p_c^{\dagger\,j \leftarrow i}(\tau),
\end{align}
where the sum over $\tau$ is over the discrete replica spin space $\{\pm 1\}^n$.

Because all layers $\alpha$ belonging to the same block $b$ receive identical
incoming fields, the contributions of the $L = M/B$ layers in block $b$ at
site $i$ factorize. Performing the sum over the replicated spins
$\{x_{i\alpha}\}$ therefore yields
\begin{align}
\sum_{\{x_{i\alpha}\}_{\alpha \in B_b}}
\prod_{\alpha \in B_b}
\prod_{k \in \partial i}
F_b^{k\leftarrow i}(x_{i\alpha})
&=
\left[
\sum_{x \in \{\pm 1\}^n}
\prod_{k \in \partial i}
F_b^{k\leftarrow i}(x)
\right]^L .
\end{align}

After collecting all such contributions over sites and blocks, we obtain



\begin{align}
\mathbb{E}_{C}[Z_M^{n}]
&\propto
\int\!\prod_{(i,j)}\prod_{b=1}^B \mathcal{D}p_b^{\,i \to j}\,\mathcal{D}p_b^{\dagger\,j \leftarrow i}\; \nonumber \\
&\exp\!\big[ -M\,S[p,p^\dagger] \big] \exp\!\big[ O(N) \big],
\end{align}
with the “action”
\begin{widetext}
\begin{align}
S[p,p^\dagger] =
\sum_{(i,j)} \sum_{b,b'=1}^B Q_{bb'}\,
\langle p_b^{\,i \to j},\,U_{ij},\,p_{b'}^{\dagger\,j \leftarrow i}\rangle
- \frac{1}{B} \sum_{i=1}^N \sum_{b=1}^B
\log \sum_{x\in\{\pm 1\}^n} \;\prod_{k \in \partial i}
\left(U_{ik} Q\,p^{\dagger\,k \leftarrow i}\right)_b(x).
\end{align}
\end{widetext}

Here
\begin{flalign}
&\langle p_b^{\,i \to j},\,U_{ij},\,p_{b'}^{\dagger\,j \leftarrow i}\rangle \nonumber\\
&= \sum_{x,\tau\in\{\pm1\}^n}
p_b^{\,i \to j}(x)\,U_{ij}(x,\tau)\,p_{b'}^{\dagger\,j \leftarrow i}(\tau).
\end{flalign}

Note that we used 
\begin{align}
\exp[O(N)] &= \exp\!\bigl(MN \cdot O(1/M)\bigr) \nonumber\\
&= \bigl(e^{O(1/M)}\bigr)^{MN} \nonumber\\
&= (1 + O(1/M))^{MN}.
\end{align}

\subsubsection{Normalization}

At the replica-symmetric saddle point, stationarity enforces $p^{\dagger} = p$, and the action is minimized by real, non-negative, normalized messages. We therefore restrict to normalized probability messages $p$, absorbing all remaining scale and phase degrees of freedom into the normalization.

The functional integral is expressed in terms of the message
fields $p_b^{\,i\to j}(x)$ associated with each oriented edge and block.
We impose the natural normalization constraint
\begin{equation}
\sum_x p_b^{\,i\to j}(x)=1,
\end{equation}
so that each $p_b^{\,i\to j}$ represents a probability distribution over the
replicated spin variable.

\subsubsection{Replica limit}

To obtain the free-energy functional we expand the replicated action
$S[p]$ in powers of $n$ and retain only the term linear in $n$, since
\begin{align}
\lim_{n\to 0} \frac{1}{n} \log \mathbb{E}_{C}\!\left[ Z_{M}^{\,n} \right]
= \inf_{p} S[p].
\end{align}

In this limit the replica spin vectors $x \in \{\pm 1\}^{n}$ collapse
onto a single Ising variable, and the replicated interaction kernel
\begin{align}
U_{ij}(x,\tau)
= \exp\!\left( \beta J_{ij} \sum_{a=1}^{n} x_{a}\tau_{a} \right)
\end{align}
reduces to the ordinary Ising factor
\begin{align}
U_{ij}(x,\tau) = e^{\,\beta J_{ij}x\tau}.
\end{align}

All terms depending on higher powers of $n$, such as cumulants of the replicated fields or the $(2^{n}-1)$-dimensional surface measure, vanish in the replica limit, leaving a functional integral over single-spin message distributions which defines the Bethe action $S[p]$.

\subsubsection{Bethe functional}

After performing the radial integration and collecting all prefactors, the replicated, $C$–averaged partition function can be written as
\begin{equation}\label{eq:betheaction}
\mathbb{E}_{C}[Z_M^{n}] \propto
\int d\mu[p] \;
\exp\!\big[ -M\,S[p] \big] \exp\!\big[ O(N) \big],
\end{equation}
where $|\mathcal{E}|$ is the number of edges in the base graph.

For each oriented edge $(i\to j)$ and block $b$, let
$p_b^{\,i\to j}(x_i)$ be the probability message over $x_i$
sent from $i$ to $j$ in block $b$.
We define its propagated field as
\begin{equation}
\hat{p}_b^{\,i\to j}(x_j)
= \sum_{c=1}^B Q_{bc} \sum_{x_i}
U_{ij}(x_j,x_i)\, p_c^{\,i\to j}(x_i),
\label{eq:hatp-def}
\end{equation}
which is the message after applying the block–mixing $Q$ and edge coupling $U_{ij}$.
The $\hat{p}$ fields are what enter multiplicatively at the receiving node in BP updates.

With this notation, the edge contribution is
\begin{equation}
Z_{ij} = \sum_{b=1}^B \sum_{x_i} p_b^{\,i\to j}(x_i)\,
\hat{p}_b^{\,j\to i}(x_i),
\end{equation}
where $\hat{p}_b^{\,j\to i}$ is the propagated field from $j$ to $i$.

The site contribution for node $i$ is
\begin{equation}
Z_{i b} = \sum_{x_i} \prod_{k\in\partial i} \hat{p}_b^{\,k\to i}(x_i).
\end{equation}
i.e. the product of incoming propagated fields from all neighbors $k$.

The large–$M$ effective action is
\begin{align}\label{eq:BetheActionp}
S[p]
= \sum_{(i,j)\in\mathcal{E}}
\log Z_{ij}
-\frac{1}{B}\sum_{i=1}^{N}\sum_{b=1}^B \log Z_{i b},
\end{align}
with $Z_{ij}$ and $Z_{i b}$ defined above and $\hat{p}$ related to $p$ by \eqref{eq:hatp-def}.

The functional integration is over normalized messages:
\begin{align}
d\mu[p] &=
\prod_{(i \to j)\in\mathcal{E}} \;\prod_{b=1}^{B}
\left[
\prod_{x} dp_b^{\,i \to j}(x) \;
\delta\!\left( \sum_{x} p_b^{\,i \to j}(x) - 1 \right)
\right].
\end{align}

For $B=1$ (only one block), $Q_{11}=1$ and the block index drops.
We have
\begin{equation} \label{eq:def-hatp-rs}
\hat{p}^{\,i\to j}(x_j) = \sum_{x_i} U_{ij}(x_j,x_i)\,p^{\,i\to j}(x_i),
\end{equation}
and the action reduces to the standard replica–symmetric Bethe free–energy functional:
\begin{widetext}
\begin{align}
S[p]
&=\sum_{(i,j)} \log 
\left[ \sum_{x_i} p^{\,i \to j}(x_i) \, \hat{p}^{\,j \to i}(x_i) \right]
-\sum_{i=1}^{N} \log 
\left[ \sum_{x_i} \prod_{k \in \partial i} \hat{p}^{\,k \to i}(x_i) \right].
\end{align}
\end{widetext}

\subsection{Belief propagation with mixing\label{sec:BPmixing}}

In this section, we apply the saddle-mode approximation at large $M$ to the Bethe functional of the M-layer (see eq. (\ref{eq:betheaction})) to obtain the message-passing equations as a function of the mixing kernel $Q$.

\subsubsection{Replica-symmetric saddle point \label{sec:saddleNode}}

Varying $Z_{ij}$ w.r.t.\ $p_b^{\,i \to j}(x)$ yields

\begin{equation}
\frac{\partial \log Z_{ij}}{\partial p_b^{\,i \to j}(x)}
= \frac{\hat{p}_b^{\,j \to i}(x)}{Z_{ij}}.
\end{equation}

For a node $j$, define
\begin{equation}
\Psi_{j,b}(x_j) = \prod_{k \in \partial j} \hat{p}_b^{\,k \to j}(x_j),
\qquad
Z_j(b) = \sum_{x_j} \Psi_{j,b}(x_j).
\end{equation}
If $k=i$, the message $p_c^{\,i \to j}$ contributes to $\Psi_{j,b}$ through the factor
$\hat{p}_b^{\,i \to j}$, so
\begin{align}
\frac{\partial \log Z_j(b)}{\partial p_c^{\,i \to j}(\tau)}
&= \frac{1}{Z_j(b)}
\sum_{x_j} \frac{\Psi_{j,b}(x_j)}{\hat{p}_b^{\,i \to j}(x_j)}
\frac{\partial \hat{p}_b^{\,i \to j}(x_j)}{\partial p_c^{\,i \to j}(\tau)} \\
&= \frac{1}{Z_j(b)}
\sum_{x_j} \frac{\Psi_{j,b}(x_j)}{\hat{p}_b^{\,i \to j}(x_j)}
\,Q_{bc}\,U_{ji}(x_j,\tau).
\end{align}

Imposing normalization with Lagrange multipliers $\lambda_b^{\,i \to j}$,
the stationarity equation $\delta S/\delta p_b^{\,i \to j}=0$ becomes
\begin{align}
0 &=
\frac{\hat{p}_b^{\,j \to i}(x_i)}{Z_{ij}} \nonumber \\
&-\frac{1}{B} \sum_{b'} Q_{b'b}
\frac{1}{Z_j(b')} \sum_{x_j}
\frac{\Psi_{j,b'}(x_j)}{\hat{p}_{b'}^{\,i \to j}(x_j)}
\,U_{ji}(x_j,x_i)
- \lambda_b^{\,i \to j}.
\end{align}
The same equation holds for the reverse orientation $(j\to i)$ with $i$ and $j$ swapped.

At the RS saddle point, the above condition is equivalent to the BP update:
for each oriented edge $(i\to j)$,
\begin{equation}
p_b^{\,i \to j}(x_i)
= \frac{1}{Z^b_{i \to j}}
\prod_{k \in \partial i \setminus j} \hat{p}_b^{\,k \to i}(x_i),
\label{eq:bp-update-rs}
\end{equation}
where the propagated fields $\hat{p}^{\,k\to i}$ are computed from the
incoming messages $p^{\,k\to i}$ via \eqref{eq:hatp-def}.
The normalization $Z^b_{i \to j}$ ensures $\sum_{x_i} p_b^{\,i \to j}(x_i)=1$.

Equation \eqref{eq:bp-update-rs} is exactly the multi–block generalization of
the standard BP update: the outgoing message $p^{\,i\to j}$ is the product
of the propagated incoming messages $\hat{p}^{\,k\to i}$ from all
neighbors $k\neq j$. For $B=1$ the block index drops, $Q_{11}=1$, and
\eqref{eq:bp-update-rs} reduces to the usual scalar BP rule.





\subsubsection{Natural parametrization\label{sec:BPfield}}

In the Ising case, each BP message \(p_{b}^{i \to j}(x_i)\) is a probability distribution over a binary variable \(x_i \in \{\pm 1\}\). At the saddle point of the Bethe functional, the stationarity conditions enforce that each message be strictly positive and normalized:
\begin{align}
p_{b}^{i \to j}(x_i) > 0, 
\qquad 
\sum_{x_i} p_{b}^{i \to j}(x_i) = 1.
\end{align}
Thus the space of messages is the interior of a 2-dimensional probability simplex. For binary variables, this space is one-dimensional: any probability distribution \((p(+1), p(-1))\) is fully determined by its magnetization
\begin{align}
m_{b}^{i \to j} = p_{b}^{i \to j}(+1) - p_{b}^{i \to j}(-1).
\end{align}
It is therefore natural to use an exponential-family parametrization. The unique canonical representation for a binary distribution with magnetization \(m\) is
\begin{align}
p(x) = \frac{1 + m x}{2}
= \frac{e^{\beta h x}}{2 \cosh(\beta h)},
\end{align}
where the “cavity field’’ \(h\) is the natural parameter (the Bernoulli message distributions belong to an exponential family), related to the magnetization by
\begin{align}
m = \tanh(\beta h), 
\qquad 
h = \frac{1}{\beta} \operatorname{artanh}(m).
\end{align}

In particular, we have 
\begin{align}
p_b^{i \to j}(x_i) = \frac{e^{\beta h_b^{i \to j} \, x_i}}{2 \cosh(\beta h_b^{i \to j})},
\end{align}
where $h_b^{i \to j}$ is the local cavity field. 
The corresponding local magnetization is the mean spin
\begin{align}
m_b^{i \to j} = \mathbb{E}[x_i] = \tanh(\beta h_b^{i \to j}).
\end{align}
Propagation through edge \((k,i)\) with block mixing \(Q\) yields a propagated factor
\begin{align} \label{eq:phat}
\hat p_b^{\,k\to i}(x_i)=C_{ik,b}\,e^{\beta u_b^{\,k\to i}x_i},
\end{align}
\noindent with 
\begin{align}
\tanh(\beta u_b^{\,k\to i})
=\tanh(\beta J_{ik})\sum_{c}Q_{bc}\,\tanh(\beta h_c^{\,k\to i}),
\end{align}
\noindent and \(C_{ik,b}=\cosh(\beta J_{ik})/\cosh(\beta u_b^{\,k\to i})\).

Inserting this parametrization into the Bethe functional and enforcing stationarity with respect to variations of \(p_{b}^{i \to j}\) yields the BP saddle-point equation
\begin{align}
p_{b}^{i \to j}(x_i) \propto \prod_{k \in \partial i \setminus j} 
\hat{p}_{\,b}^{k \to i}(x_i),
\end{align} which in the exponential-family form becomes simply an additive update in the field variables:
\begin{align}
\beta h_{b}^{i \to j}
= \sum_{k \in \partial i \setminus j}
\operatorname{atanh}\!\left[
\tanh(\beta J_{ik})\, (Q m^{k \to i})_b
\right],
\end{align}
or equivalently,
\begin{align}
h_{b}^{i \to j}
= h_i
+ \sum_{k \in \partial i \setminus j}
u_{b}^{k \to i},
\end{align}
where \(h_i\) is the local field (if present) and \(u_{b}^{k \to i}\) is the effective field sent from neighbor \(k\) to \(i\) through block \(b\) (given explicitly in eq. (\ref{eq:h-bp})). In this parametrization the BP update becomes linear in the fields even though it is nonlinear in the underlying probabilities. This makes the field representation the natural choice for both analytical derivations and numerical algorithms.

\subsection{Dynamics of BP with mixing}

\subsubsection{Momentum in BP\label{sec:momentumBP}}

If $Q$ is normal but not symmetric, for instance a circulant ring kernel with drift (see Fig. \ref{fig:Ising_summary}), $Q$ can be diagonalized by a discrete Fourier transform $Q = V \Phi V^*$ with $V_{\alpha k} = \frac{1}{\sqrt{B}} \, e^{-2\pi i \frac{(\alpha - 1)(k - 1)}{B}}$ and complex eigenvalues $\phi_k = \rho_k e^{i\omega_k}$. Eliminating the phase gives a two-step recurrence identical to the heavy-ball or momentum form for each joint mode of the lifted system given as (when $\eta = 1$):
\begin{align}
\delta_{j,k,t+1} = (1 - \lambda_{j,k} + \beta_{j,k})\, \delta_{j,k,t}
- \beta_{j,k}\, \delta_{j,k,t-1},
\end{align}
with $\beta_{j,k} = |\nu_j \phi_k|^2$ and $\lambda_{j,k} = 1 + |\nu_j \phi_k|^2 - 2\, \Re(\nu_j \phi_k)$ where $\nu_j$ are the eigenvalues of $W \, K$. The drift angle $\omega_k$ introduces oscillatory ``momentum'', while the width $\sigma$ of $Q$ controls the damping $\rho_k$.  Mixing therefore acts like a distributed momentum term that accelerates synchronization while preserving the same collapse condition $|\nu_j| \rho_k < 1$.

\subsubsection{Projection on homogenous\label{sec:projhomo}}

In this section, BP iteration of eq.~(\ref{eq:BPG}) are projected onto the block-uniform manifold.
\\
\\
We define the projection onto the block-uniform manifold $\Pi = I \otimes \Pi_B$ with $\Pi_B = \frac{1_B 1_B^\top}{B} = \frac{1}{B} 1_B 1_B^\top$ and
$\Pi^\perp = I - \Pi$ its orthogonal subspace (the fluctuation subspace). Given $h = \bar{h} + \delta$, we thus have $\Pi h = \bar{h}$ and $\Pi \delta = 0$.
\\
\\
We expand eq.~(\ref{eq:BPG}) in $\delta$ as follows:
\begin{align}
\bar{h}^+ = G(\bar{h}) + DG|_{\bar{h}} \delta + D^2 G|_{\bar{h}} [\delta,\delta] + O(|\delta|^3).
\end{align}
\\
\\
First, consider the linear part in $DG|_{\bar{h}} \delta$. If $Q$ is doubly stochastic, then
\begin{align}
\Pi D_G|_{\bar{h}} \Pi^\perp &= (1 - \eta)\Pi \Pi^\perp + \eta \Pi D_F|_{\bar{h}} \Pi^\perp \nonumber \\
&= \eta (W K) \otimes (\Pi_B Q \Pi_B^\perp) = 0.
\end{align}
That is, fluctuations inject no linear bias into the uniform manifold.
\\
\\
Next, we calculate the second order term $D^2 G|_{\bar{h}} [\delta,\delta] = \eta D^2 F|_{\bar{h}} [\delta,\delta]$. Recall that $F(h) = \sum_{k \in \partial i \setminus j} u_b^{k \to i}$ (see eq.~(\ref{eq:h-bp})) with $u_b^{e}$ given as follows for each edge $e$ and block $b$:
\begin{align}
\tanh(\beta u_b^e) = \lambda_e \sum_c Q_{bc} \tanh(\beta h^e_c).
\end{align}
with $\lambda^e = \tanh(\beta J_{ik})$. that is $u_b^e = g\big((Q m(h^e))_b\big)$ with
\begin{align}
g(x) = \beta^{-1} \operatorname{artanh}(\lambda^e x), \quad m(h) = \tanh(\beta h).
\end{align}
A second–order expansion at $\bar{h}$ yields
\begin{widetext}
\begin{align}
D^2 u_b^e|_{\bar{h}}[\delta^e, \delta^e]
= \underbrace{\beta^2 s_e^2 g''(m_e)}_{\Gamma_e}
((Q \delta_e)_b)^2
+ \underbrace{\beta^2 g'(m_e)(-2 m_e s_e)}_{-\Gamma'_e}
\sum_c Q_{bc} (\delta^e_{c})^2.
\end{align}
\end{widetext}
with $D^{2}u^e_{b} |_{\bar{h}} \, [\delta_e, \delta_e] = \sum_{c,c'} \frac{\partial^{2} u^e_{b}}{\partial h^{e}_{c} \, \partial h^{e}_{c'}} |_{\bar{h}} \delta^{e}_{c} \delta^{e}_{c'}$ and 
\begin{align}
g'(m_e) = \frac{\lambda_e}{ \beta (1 - \lambda_e^2 m_e^2)}, \quad
g''(m_e) = \frac{2 \lambda_e^3 m_e}{ \beta (1 - \lambda_e^2 m_e^2)^2}.
\end{align}
given
\begin{align}
m_e = \tanh (\beta \bar{h}^e), \quad s_e = \sech^2(\beta \bar{h}^e).
\end{align}
We have:
\begin{align}
\Gamma_e = \frac{2 \beta \lambda_e^3 m_e s_e^2}{ (1 - \lambda_e^2 m_e^2)^2}, \quad
\Gamma'_e = \frac{2 \beta \lambda_e m_e s_e}{ 1 - \lambda_e^2 m_e^2}.
\end{align}

Note that we have $(h^+)^e_b = F^e_b(h) = (W u_b(h))^e$. Averaging over blocks with $\Pi_B$ gives, for each edge $e$,
\begin{align}
(\Pi D^2 F|_{\bar{h}}[\delta, \delta])^e
= \frac{1}{B} \big(
\Gamma_e(Q) \|Q \delta^e\|_2^2
- \Gamma_e(I) \|\delta^e\|_2^2
\big).
\end{align}
Since $D^2 G = \eta D^2 F$, the quadratic forcing on the block average is
\begin{align}
\xi = \frac{\eta}{2B} W \big(
\Gamma q_Q(\delta) - \Gamma' q_I(\delta)
\big),
\end{align}
where $\Gamma = \operatorname{diag}(\Gamma_e)$ and $\Gamma' = \operatorname{diag}(\Gamma'_e)$ over oriented edges with $q_Q(\delta) = \|Q \delta^e\|_2^2$ and $q_I(\delta) = \|\delta^e\|_2^2$.
\\
\\
In summary:
\begin{align}
\bar{h}^+ = G(\bar{h}) + \xi + O(\|\delta\|^3),
\end{align}
\begin{align}
\delta^+ = \Pi^\perp D_G|_{\bar{h}} \Pi^\perp \, \delta + O(\|\delta\|^2).
\end{align}

\subsection{1-RSB formalism \label{sec:1RSB}}

\subsubsection{1-RSB ansatz}

For a fixed oriented edge $(i\to j)$ and block $b$, the replicated message
$p_b^{\,i\to j}(x^1,\ldots,x^n)$ is symmetric in replicas. In
1‑RSB with Parisi parameter $x\in(0,1]$, replicas are partitioned into $n/x$
groups of size $x$ and are exchangeable within each group. By de Finetti,
within a group spins are conditionally i.i.d.\ Bernoulli with a latent
``spin‑up'' probability $\pi\in(0,1)$ with
\begin{align}
\mathbb{P}(x=+1\mid \pi)=\pi, \qquad
\mathbb{P}(x=-1\mid \pi)=1-\pi,
\end{align}
let $\mu_b$ be a probability measure on $\pi$ (the intra‑state field
distribution for block $b$), and let $P_b^{\,i\to j}$ be a law on such measures
(a ``survey'' over pure states), then
\begin{flalign}
&p_b^{\,i\to j}(\{x^a\}) \nonumber \\
&=\int P_b^{\,i\to j}(d\mu_b)\,
\Bigg[\int \mu_b(d\pi)\;\prod_{a=1}^{x}
\pi^{\frac{1+x^a}{2}}(1-\pi)^{\frac{1-x^a}{2}}\Bigg]^{\!n/x}.
\end{flalign}

Write $m=2\pi-1\in(-1,1)$ and $h=\beta^{-1}\operatorname{artanh} m$ so that
$\pi=(1+\tanh\beta h)/2$ and
\begin{align}
\pi^{\frac{1+x}{2}}(1-\pi)^{\frac{1-x}{2}}
=\frac{e^{\beta hx}}{2\cosh(\beta h)}.
\end{align}
After the $\pi \mapsto h$ change of variables, 
$P_b^{i \to j}$ becomes a measure on measures $\nu_b$ over cavity fields $h \in \mathbb{R}$.
The ansatz becomes the standard
mixture of RS group factors in $h$:
\begin{align}
p_b^{\,i\to j}(\{x^a\})
=\int P_b^{\,i\to j}(d\nu_b)\,
\Bigg[\int \nu_b(dh)\;\prod_{a=1}^{x}
\frac{e^{\beta h x^a}}{2\cosh(\beta h)}\Bigg]^{\!n/x}\!,
\end{align}
i.e. “draw a state $\nu_b$, then draw one field $h$ per replica‑group, and make
the $x$ spins in the group i.i.d.\ with law $e^{\beta hx}/(2\cosh \beta h)$.”

Note that, because the block-mixing $Q$ couples the $B$ components at propagation, the object to consider is the joint law on $h = (h_1, \dots, h_B)$. For readability we present component-wise formulas, with the understanding that the input to each block-$b$ update depends on the vector of incoming fields $\{h_c\}_{c=1}^B$ through the $Q$-mixing.
\\
\\
In summary, exchangeability implies that across pure (Bethe) states, the RS message on each oriented edge becomes a random variable; the 1-RSB order parameter is a probability law over messages (a law over block-field vectors $\boldsymbol{h} = (h_{1},\dots,h_{B})$).

\subsubsection{1-RSB factor graph of messages}

\begin{figure*}[t]
    \centering
    \includegraphics[width=1.0\textwidth]{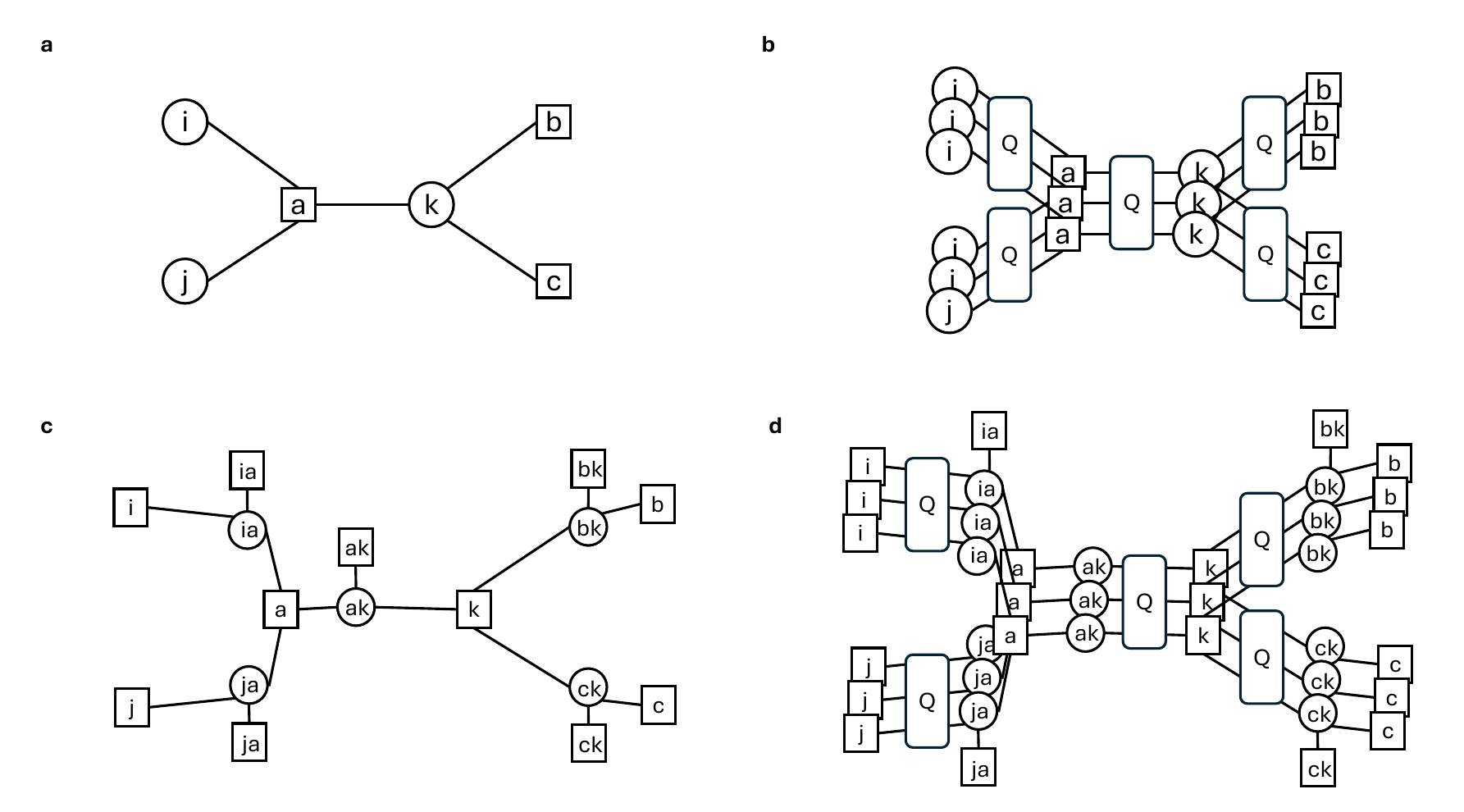}
    \caption{
    a) Base factor graph with variables (circles) and factors (squares).  
b) $M$-layer lift: the entire factor graph is copied $M$ times and corresponding variable–factor incidences are rewired across layers according to permutations drawn from the mixing matrix $Q$.  
c) 1-RSB representation of the base graph: each message $(i,a)$ becomes a variable. Variables $i$, factors $a$, and each edge $(i,a)$ is expanded into factors enforcing the BP iterative equations and 1-RSB reweighting.  
d) 1-RSB representation of the $M$-layer lift: the construction in (c) is replicated $M$ times and rewired across layers using the same $Q$-distributed permutations as in (b).
}
    \label{fig:1RSBschema}
\end{figure*}

When interpreting RS messages a random variables, we can now interpret the probability of messages as an auxiliary probabilistic graphical model in which messages are variables and RS BP equations define factors (see Fig. \ref{fig:1RSBschema}).
\\
\\
\\
\\
We define an auxiliary graphical model whose variables are the RS messages on each edge, \(\{ p^{i \to j}, \,\hat p^{j \to i} \}\) and impose the mixed-BP constraints 
\begin{align} 
p = f(\hat p), \qquad \hat p = \hat f(p;Q)
\end{align} 
as hard factors where $f$ and $\hat f$ are given by eqs. (\ref{eq:BPinp}) and (\ref{eq:hatp-def-main}), and weight each message configuration by the Bethe free energy $F$ (which is $F[p]=-S[p]$ with $S$ the Bethe action) of 
the \(M\)-layer (see eq. (\ref{eq:action-main})) raised to \(x\): 
\begin{align} 
W_{x}[p] &= e^{x S[p]},
\end{align} 
\noindent with 
\begin{align}
S[p]=\sum_{(i,j)\in E}\log Z_{ij}
-\frac{1}{B}\sum_{i\in V}\sum_{b=1}^B\log Z_{ib},
\end{align} 
with \(Z_{ij}\), \(Z_{ib}\), and \(\hat p\) defined in
Eqs.~(\ref{eq:Zij-main})--(\ref{eq:hatp-def-main}).
Applying BP on this auxiliary model yields the 1-RSB recursive distributional
equations, as in the standard cavity construction but with block mixing entering
through \(\hat f\).

For the purpose of the 1-RSB reweighting, it is convenient to rewrite the Bethe
free energy \(F[p]=-S[p]\) in a directed, gauge-dependent form.
We therefore decompose \(F[p]\) as
\begin{equation}
F[p]
=
\sum_{(ij)\in E} F_{ij}
+
\sum_{i\in V} F_i
-
\sum_{(i\to j)\in\vec E} F_{i,(ij)},
\label{eq:BetheGaugeDecomp}
\end{equation}
where the undirected edge and site contributions are defined as
\begin{align} 
F_{ij} &= \log Z_{ij},\\
F_i &= \frac{1}{B}\sum_{b=1}^B \log Z_{ib},
\end{align} 
and \(F_{i,(ij)}\) denotes a directed cavity contribution.
The consistency condition
\begin{align} 
\sum_{(i\to j)\in\vec E} F_{i,(ij)} = 2\sum_{(ij)\in E} F_{ij}
\end{align} 
ensures that Eq.~(\ref{eq:BetheGaugeDecomp}) is equivalent to the symmetric form of the Bethe free energy.

To make the directed contribution explicit, we introduce the block-resolved
cavity normalizers
\begin{equation}
z_{\,b}^{i\to j}
=
\sum_{x_i=\pm1}
\prod_{k\in\partial i\setminus j}
\hat p_b^{\,k\to i}(x_i),
\end{equation}
which normalize the RS BP updates
\(
p_b^{\,i\to j}(x_i)\propto\prod_{k\neq j}\hat p_b^{\,k\to i}(x_i)
\).
The directed cavity free-energy term is then defined as the block average
\begin{equation}
F_{i,(ij)} = \frac{1}{B}\sum_{b=1}^B \log z_{\,b}^{i\to j}.
\end{equation}

It is also convenient to introduce the associated variable- and factor-side
normalization factors\cite{mezard2009information}
\begin{equation}
z_{i\to j} = e^{F_i - F_{i,(ij)}},
\qquad
\hat z_{j\to i} = e^{F_{ij} - F_{i,(ij)}}.
\end{equation}
These quantities correspond, respectively, to the normalization of the
variable-to-factor and factor-to-variable BP updates.

Because the Bethe free energy is invariant under local gauge transformations
that redistribute contributions between site, edge, and directed cavity terms,
we may choose a reference-orientation gauge in which
\begin{equation}
F_{i,(ij)} = F_{ij}.
\end{equation}
In this gauge, the factor-to-variable normalization reduces to
\begin{equation}
\hat z_{j\to i} = 1,
\end{equation}
and all reweighting is carried by the variable-side cavity normalizers
\(z_{i\to j}\). This choice is particularly convenient for obtaining a
neighbor-separable reweighting in the zero-temperature limit.

\subsubsection{Change gauge}

We further apply a Bethe gauge:
\begin{align} 
F_{i} &\;\mapsto\; \widetilde F_{i} = F_{i} - H_{i}, \\
F_{i,(ij)} &\;\mapsto\; \widetilde F_{i,(ij)} = F_{i,(ij)} - H_{i,(ij)}, \\
F_{ij} &\;\mapsto\; \widetilde F_{ij} = F_{ij} + \Delta_{ij},
\end{align} 
\noindent which keeps \(F\) unchanged when we choose
\begin{align} 
\sum_{\{ij\}} \Delta_{ij}
\;=\;
\sum_{(i\to j)} H_{i,(ij)}
\;-\;
\sum_{i} H_{i}.
\end{align} 

We set in particular:
\begin{align} 
\widetilde F_{i}
&\;=\;
F_{i}
-
\frac{1}{B}\sum_{b}\log\!\bigl(2\cosh(\beta S_{b})\bigr), \\
\widetilde F_{i,(ij)}
&\;=\;
F_{i,(ij)}
-
\frac{1}{B}\sum_{b}\log\!\bigl(2\cosh(\beta S_{b \setminus j})\bigr),
\end{align} 

\noindent with $S_b=\sum_{k \in \partial i} u_b^{k \to i}$ and $S_{b\setminus j}=\sum_{k \in \partial i \setminus j} u_b^{k \to i}$. This choice is motivated by the fact that this will lead to neighbor-separable reweighting at zero temperature allowing to sample each neighbor of a site independently.

Recall that $\hat p^{\,b}_{k\to i}(x_i)=C_{ik,b}\,e^{\beta u^{\,b}_{k\to i}x_i}$
(see Eq.~(\ref{eq:phat})). Therefore, the block-resolved site partition factor is
\begin{align}
Z_{ib}
&=\sum_{x_i=\pm1}\prod_{k\in\partial i}\hat p_b^{\,k\to i}(x_i)
=2\left(\prod_{k\in\partial i}C_{ik,b}\right)\cosh(\beta S_b).
\end{align}

Since the Bethe free energy contains the block-averaged site contribution
$F_i=\frac{1}{B}\sum_b\log Z_{ib}$, the difference between the site and cavity
free energies entering the 1-RSB reweighting can be written as
\begin{align}
&F_i-F_{i,(ij)} = \nonumber \\
&
\frac{1}{B}\sum_{b}
\Bigl[
\log\cosh(\beta S_b)
-
\log\cosh(\beta S_{b\setminus j})
\Bigr]
+\text{Cst},
\end{align}
where the constant terms independent of the messages have been dropped.

With the above gauge choice, the corresponding 1-RSB reweighting factor reads
\begin{align}
w(h,J) &= e^{x(\widetilde F_i-\widetilde F_{i,(ij))})} \nonumber\\
&=
e^{\!\left\{
-\frac{x}{B}\sum_{b}
\bigl[
\log\cosh(\beta S_b)
-
\log\cosh(\beta S_{b\setminus j})
\bigr]
\right\}},
\end{align}
which is neighbor-separable and leads to a factorized reweighting in the
zero-temperature limit.

\subsubsection{Zero-temperature limit}

At zero temperature ($\beta \rightarrow \infty$), we get the following BP iterative equations in the cavity field $h$:
\begin{align}
(h_{\,b}^{i\to j})^+
=
h_i
+
\sum_{k\in\partial i\setminus j}
u_{\,b}^{k\to i}.
\end{align}
\noindent with 
\begin{align}
u_{\,b}^{k\to i}
=
\frac{1}{2}
\bigl(
\lvert J_{ik} + H_{\,b}^{k\to i}\rvert
-
\lvert J_{ik} - H_{\,b}^{k\to i}\rvert
\bigr)
\end{align}
\noindent and $H$ with $H_{\,b}^{k\to i} = (Q h^{k\to i})_{b}$ is the mixed field.
\\
\\
Moreover, the variable side local normalizers, noted $w(h,J)$, can be rewritten as follows using $\log\cosh(\beta t)=\beta |t| + o(\beta)$:

\begin{align}
w(h,J) = e^{\,-x(\widetilde F_{i}-\widetilde F_{i,(ij)})}
\;\propto\;
\prod_{k\in\partial i\setminus j}
\exp\!\left\{ \frac{y}{B}\sum_{b=1}^{B}
a_{b}\!\left(h_{k\to i},\,J_{ik}\right)
\right\}
\end{align}
where \(y=\beta x\) and $a_{b}(h,J) = \frac{ \bigl|J + (Qh)_{b}\bigr| + \bigl|J -  (Qh)_{b}\bigr| }{2}$.

The RDE equation is then given as follows:
\begin{widetext}
\begin{align}
P^{\mathrm{out}}_{i\to j}(dh^{i\to j})
\;\propto\;
\int
\Biggl[
\prod_{k\in\partial i\setminus j}
P_{k\to i}\bigl(d h^{k\to i}\bigr)
\Biggr]\,
K_{i\to j}\!\left(\{ h^{k\to i}\}\right)\,
\delta^B\!\left(
h^{i\to j}
-
\left[ \sum_{k \in \partial i \setminus j} u(h^{k\to i},J_{ik}) \right]
\right)
\end{align}
\end{widetext}
\noindent where $K_{i\to j} \propto \prod_{k\in\partial i\setminus j}
w\!\left(h^{k\to i}\,;\,J_{ik}\right)$ and $u_{b}(h,J) = \frac{ \bigl|J + (Qh)_{b}\bigr| - \bigl|J -  (Qh)_{b}\bigr| }{2}$. We have used the notation $\delta^B$ for the vector delta (the Kronecker delta product over $b$).

When $B=1$, we recover
\begin{align}
a(h,J)&=\frac{1}{2}\bigl(|J+h|+|J-h|\bigr),\\
u(h,J)&=\frac{1}{2}\bigl(|J+h|-|J-h|\bigr),
\end{align}
and therefore
\begin{align}
w &\propto \prod_{k\in\partial i\setminus j}
\exp\!\bigl[y\,a(h^{k\to i},J_{ik})\bigr],\\
h_{i\to j}
&=
\sum_{k\in\partial i\setminus j}
u(h^{k\to i},J_{ik}),
\end{align}

which is the standard \(T=0\) cavity theory\cite{mezard2003cavity}.

\subsection{Construction of permutations\label{sec:generation}}

We detail here how the $M$-layer lift is numerically constructed in practice for the Ising model. Each edge in the base coupling matrix \(J\) is replicated across \(M\) layers, with inter-layer wiring sampled from a permanental distribution derived from the mixing matrix \(Q\). 
\\
\\
Given a non-negative matrix \(Q\), we first compute a doubly stochastic normalization
\begin{align}
\tilde Q = \operatorname{diag}(r)\, Q\, \operatorname{diag}(c),
\end{align}
obtained by iterative Sinkhorn rescaling (obtained by alternating row and column normalization) in the log-domain to machine precision. Sinkhorn balancing rescales $Q$ to be approximately doubly stochastic, removing marginal inhomogeneities so that the subsequent permanental weighting is not dominated by trivial row/column scale differences.
\\
\\
The goal is then to draw a permutation \(\pi \in S_M\) with probability proportional to the permanent of the submatrix selected by \(\pi\):
\begin{align}
P(\pi) \propto \prod_{a=1}^{M} \tilde Q_{a,\pi(a)} = \exp (\sum_a \log \tilde{Q}_{a,\pi(a)}) .
\end{align}

To generate biased inter-layer permutations, we use an approximate perturb-and-MAP sampler based on entrywise Gumbel perturbations\cite{papandreou2011perturb,tarlow2012randomized} followed by a Hungarian assignment solve. This procedure provides a tractable heuristic for sampling permutations biased by the permanental weight induced by $Q$, with computational complexity $O(M^3)$.




For each nonzero coupling $|J_{ij}| > 0$ with $i<j$, we draw a pair of layer permutations
$\pi_{(i,j),i}$ and $\pi_{(i,j),j}$ from the distribution $P(\pi)$. These define the relative permutation
\begin{align}
\sigma_{ij} \;=\; \pi_{(i,j),j}\circ \pi_{(i,j),i}^{-1},
\end{align}
which determines how the interaction between spins $i$ and $j$ is wired across the
$M$ layers. In the lifted adjacency, node $(i,\alpha)$ in layer $\alpha$ is connected to
node $(j,\sigma_{ij}(\alpha))$ in the corresponding permuted layer. In the symmetric case, we have $\sigma_{ji}=\sigma_{ij}^{-1}$ ensuring undirected interactions in the lifted graph. Using the flattened
site index $(\alpha,i)\mapsto (\alpha-1) n + i$, this produces a sparse $nM\times nM$
matrix
\begin{align}
J^{(M)}_{(\alpha,i),(\beta,j)} \;=\;
\begin{cases}
J_{ij}, & \beta = \sigma_{ij}(\alpha),\\[4pt]
0,      & \text{otherwise},
\end{cases}
\end{align}
so that each base edge $(i,j)$ lifts to $M$ cross-layer edges consistent with the
permutation structure. The resulting block pattern of $J^{(M)}$ implements the
couplings appearing in the lifted Hamiltonian~\eqref{eq:IsingH}.

\subsection{Empirical benchmark\label{sec:benchmark}}

\subsubsection{Ising problems}

The benchmark results shown in Fig. \ref{fig:benchmark} is constructed as follows (see table \ref{table:problems}). We construct regular random graph (RRG) of degree $d=3$ with binary random interactions $J_{ij} = \pm 1$, binary Sherrington-Kirkpatrick instances with $J_{ij} = \pm 1$, tile planted instances\cite{perera2020chook} of dimension 2 with patterns $(p_1,p_2,p_3) = (0.2,0.5,0.1)$. The ground-state is known by construction for tile-planted instances. For other problems, the lowest energy found across all runs, parameters, and solvers is interpreted as an approximate the ground-state. Given the problem sizes considered and repetition of lowest energy found, these correspond to real ground-state with high probability.

\begin{table*}[t]
\centering
\begin{tabular}{lccc}
\hline
\textbf{Instance type} & \textbf{Degree} & \textbf{Interaction values \(J_{ij}\)} \\
\hline
Regular Random Graph (RRG) & $d = 3$ (Bethe lattice) & $J_{ij} = \pm 1$ \\[4pt]
Tile-planted (2D, $(p_1,p_2,p_3)=(0.2,0.5,0.1)$) & $d=4$ (square lattice) & 
$J_{ij} \in \{+1,\,-1,\,-2\}$ \\[4pt]
Sherrington--Kirkpatrick (SK) & $d=N$ (fully connected) & $J_{ij} = \pm 1$ \\
\hline
\end{tabular}\label{table:problems}
\end{table*}

The greedy algorithm (zero-T quench) and simulated annealing (SA) are implemented using the Glauber update eqs. (\ref{eq:glauber}). GA is iterated at zero temperature ($\beta \rightarrow \infty$) for $T$ sweeps. Each sweep updates all spins exactly once, but in a new random order drawn uniformly at random for every sweep and every replica. 

SA performs temperature-annealed single-spin Glauber updates on a possibly asymmetric coupling matrix $J$. The inverse temperature follows a linear schedule from $\beta_{\text{start}}$ to $\beta_{\text{end}}$ across the prescribed number of sweeps. The start and end temperatures are determined automatically by sampling typical uphill energy changes $\Delta E > 0$: the initial temperature is chosen to give a high target acceptance (default $0.85$), and the final temperature to give a low acceptance (default $0.02$), with fallback estimates based on column norms of $J$ when sampling is unreliable. By default, spin-flip decisions use only the incoming
local field $h_i = \sum_j J_{ij}\, x_j$ meaning the algorithm naturally supports directed (asymmetric) couplings without requiring any symmetrization of $J$.

\subsubsection{Maximum independent set \label{sec:MISappendix}}

The maximum independent set (MIS) instances are defined on a simple undirected graph $G = (V, E)$ with binary variables $n_i \in \{0,1\}$ indicating whether vertex $i$ is included in the independent set. The hard
constraint is
\begin{align}
n_i n_j = 0 \quad \text{for all } (i,j)\in E,
\end{align}
i.e., no edge may have both endpoints simultaneously occupied. The MIS density is
\begin{align}
\rho = \frac{1}{|V|} \sum_{i\in V} n_i.
\end{align}

A standard quadratic Ising formulation introduces spin variables
$s_i \in \{-1,+1\}$ with
\begin{align}
n_i = \frac{1 + s_i}{2}.
\end{align}
A common penalty-based MIS Hamiltonian is
\begin{align}
H(s)
= -\sum_i \mu_i\, n_i
\;+\;
\lambda \sum_{(i,j)\in E} n_i n_j.
\end{align}
where $\lambda > \max_i \mu_i$ enforces the independence constraint.
\\
\\
The $\mu$SA\cite{angelini2019monte} algorithm is a zero-temperature single spin dynamics for the maximum independent set in which a chemical-potential-like parameter $\mu_i = \mu$, $\forall i$, is slowly increased from $\mu=0$. The configuration starts empty ($n_i=0$ for all vertices) on a fixed graph, and each Monte Carlo sweep visits every vertex exactly once in a random permutation order. When a vertex $i$ is currently empty, it is occupied ($n_i \leftarrow 1$) if and only if all its neighbors are empty (no constraint violation); when it is occupied, it is removed ($n_i \leftarrow 0$) with probability
$\exp(-\mu)$. After each sweep, $\mu$ is increased by a small increment $\Delta\mu$, and the independent-set density $\rho(\mu) = K/N$ (with $K = \sum_i n_i$) is recorded. For replicated simulated annealing ($\mu$RSA), refer to \cite{angelini2019monte}.
\\
\\
Parallel tempering in chemical potential ($\mu$PT) extends $\mu$SA by running multiple
replicas of the system at different values of the chemical potential. We consider
$N_\mu$ replicas, each constrained to be a valid independent set, but evolving at a
distinct chemical potential
\begin{align}
\mu_i = \mu_{\max} - i\,\Delta\mu,\qquad i = 0,\ldots, N_\mu-1 .
\end{align}
Each replica independently performs standard zero-temperature Metropolis updates at its
own chemical potential: if a vertex $i$ is empty, it is occupied only if all its
neighbors are empty; if it is occupied, it is removed with probability $e^{-\mu_i}$.
A Monte Carlo sweep visits all vertices once in random order.
Every few sweeps (five, as in ~\cite{angelini2019monte}), a swap move is attempted between neighboring replicas $i$ and $i+1$. The swap exchanges their entire configurations $(n^{(i)}, n^{(i+1)})$ and is accepted with probability
\begin{align}
p_{\mathrm{swap}} = \min\Bigl(1,\,
\exp\!\bigl[(\mu_i - \mu_{i+1})(-K_i + K_{i+1})\bigr]\Bigr),
\end{align}
where $K_i = \sum_{v} n_v^{(i)}$ is the current independent-set size in replica $i$. This exchange step allows the low--$\mu$ (high-entropy) replicas to help high--$\mu$ replicas escape metastable configurations.
\\
\\
For the M-layer benchmarks, we lift each MIS instance to an M-layer factor graph and
then run $\mu$SA or $\mu$PT directly on the lifted graph, projecting back to the base
graph only for evaluation. In the M-layer $\mu$SA experiment, we first generate a random regular graph $G$ and its adjacency $J_0$, build its structured M-layer lift $G_M$ via the mixing kernel $Q$ (Gaussian-drift ring), and run zero-temperature $\mu$SA on $G_M$ with a linear schedule $\mu(t)$ from $\mu_{\min}$ to $\mu_{\max}$ over a prescribed number of sweeps. During the run we keep track of the best lifted configuration (largest number of occupied vertices). At the end, this best lifted configuration $n \in \{0,1\}^{MN}$ is projected back to the base graph by consensus + greedy: we count, for each base vertex, how many layers occupy it, sort vertices by this score, and greedily build a valid independent set on the base graph. The resulting base density $\rho$ is recorded as a function of the total number of sweeps; averaging over disorder instances and fitting $\tau(\rho)$ with a power law yields the M-layer $\mu$SA algorithmic threshold.
\\
\\
In the M-layer $\mu$PT experiment, we follow the same lifting and projection pipeline
but replace $\mu$SA with parallel tempering in chemical potential on the lifted graph.


\input{loop}

\end{document}